\documentclass[aps,prd,showpacs,superscriptaddress,preprintnumbers,nofootinbib,notitlepage]{revtex4-1}
\pdfoutput=1
\usepackage[T1]{fontenc}
\usepackage{bm,amsmath,amssymb,color,bbold,mathrsfs} 
\usepackage[pdftex]{graphicx}
\usepackage[colorlinks=true, pdfstartview=FitV, linkcolor=blue, citecolor=red, urlcolor=blue]{hyperref}

\graphicspath{{./1_Figures/}}

\DeclareMathOperator\ee{e}
\DeclareMathOperator\tr{tr}

\newcommand{\der}{\partial}
\renewcommand{\bar}[1]{\overline{#1}}
\newcommand{\calO}{\mathcal{O}}
\newcommand{\dd}{\mathrm{d}}
\newcommand{\bep}{\begin{pmatrix}} 
\newcommand{\eep}{\end{pmatrix}}

\newcommand{\SU}{\text{SU}}

\newcommand{\U}{\text{U}}
\newcommand{\1}{\mathbb{1}}

\newcommand{\ZZ}{\mathbb{Z}}
\renewcommand{\epsilon}{\varepsilon}

\def\ba#1\ea{\begin{align}#1\end{align}}

\def\mkakko#1{\left( #1 \right)}
\def\ckakko#1{\left\{ #1 \right\}}
\def\kkakko#1{\left[ #1 \right]}

\newcommand{\up}{\uparrow}
\newcommand{\down}{\downarrow}
\newcommand{\npsi}{\Psi}
\newcommand{\bk}{\mathbf{k}}
\newcommand{\B}{\text{B}}
\newcommand{\Y}{\mathscr{X}}
\newcommand{\GG}{\bar{G}}
\renewcommand{\sc}[1]{}

\begin{document}
\preprint{RIKEN-QHP-242} 
\title{Overscreened Kondo effect, (color) superconductivity 
and Shiba states in Dirac metals and quark matter
}

\author{Takuya Kanazawa}
\affiliation{iTHES Research Group and Quantum Hadron Physics Laboratory, 
RIKEN, Wako, Saitama 351-0198, Japan}
\author{Shun Uchino}
\affiliation{RIKEN Center for Emergent Matter Science, Wako, 
Saitama 351-0198, Japan}
\allowdisplaybreaks

%%%%%%%%%%%%%%%%%%%%%%%%%%%%%
%\[
%	\fbox{\quad\begin{minipage}{.73\textwidth}
%		\vspace{8pt}\tableofcontents\vspace{10pt}
%	\end{minipage}\quad}
%\]
%\newpage
%%%%%%%%%%%%%%%%%%%%%%%%%%%%%

\begin{abstract}
	We study the interplay between the Kondo effect and (color) superconductivity   
	in doped Dirac metals with magnetic impurities and in quark matter with colorful impurities. 
	We first point out that the overscreened Kondo effect arises in the normal state of these systems. 
	Next the (color) superconducting gap is incorporated as a mean field and the phase diagram 
	for a varying gap and temperature is constructed nonperturbatively. 
	A rich phase structure emerges from a competition of effects unique to a multichannel system. 
	The Kondo-screened phase is shown to disappear for a sufficiently large gap. 
	Peculiarity of quark matter due to the confining property of non-Abelian gauge fields is noted.  
	We also investigate the spectrum of sub-gap excited states, called Shiba states. 
	Based on a model calculation and physical reasoning we predict that, as the coupling of the impurity 
	to the bulk is increased, there will be more than one quantum phase 
	transition due to level crossing among overscreened states.  
\end{abstract}
\maketitle
\section{\label{sc:intro}Introduction \sc{sc:intro}}

Understanding various phenomena caused by impurities is one of the fundamental challenges in quantum many-body physics.
In condensed matter physics, an impurity is often treated as a potential term in a
Hamiltonian and causes
significant phenomena such as the Friedel oscillation and the orthogonality 
catastrophe~\cite{Mahan2000}.
These phenomena are relevant to quantum dot systems~\cite{Nazarov2009}
and $X$-ray absorption~\cite{Mahan2000},
and the essential ingredient is the existence of the Fermi sea.
If an impurity has an internal degree of freedom and the interaction with
the conduction fermions is non-Abelian, more intriguing phenomena occur.
A celebrated example is the Kondo effect \cite{Kondo01071964} signified by 
the increase of electrical resistance of alloys with dilute
magnetic impurities as decreasing temperature.    
The problem of clarifying the whole crossover of an impurity spin 
from a doublet state at high temperature to a singlet state at low temperature 
in the case of \emph{antiferromagnetic} coupling between the impurity and conduction electrons,  
has defied a straightforward perturbative solution and prompted the development 
of a variety of nonperturbative techniques, e.g., 
Anderson's poor man's scaling \cite{Anderson1970}, 
the large-$N$ slave boson approach \cite{Read-Newns1983,Coleman1984}, and 
Wilson's numerical renormalization group (NRG) \cite{RevModPhys.47.773}, 
all of which have made a substantial contribution to the foundation of a modern theoretical 
framework for heavy fermion systems 
\cite{RevModPhys.47.773,HewsonBook,Coleman2002review,KondoBook,Coleman:2015uma}. 
In recent years, interests in the Kondo 
effect were refueled by advances in nanotechnology and new discoveries 
continue to be made \cite{Kouwenhoven2001}.  

A necessary condition for the Kondo screening to occur is 
the existence of the Fermi surface. 
What happens if the Fermi surface is destroyed? This question has a long history of research: 
there are at least three distinct setups to address this question. 
First and foremost, magnetic impurities in fully gapped superconductors had attracted great 
interest \cite{RevModPhys.78.373}. Since the $s$-wave Cooper pairing induces a gap $\Delta$ 
in the fermion spectra, it competes with the Kondo effect and even eliminates it 
when $\Delta$ is sufficiently greater than $T_{\rm K}$, the Kondo temperature. 
Conversely, a finite concentration of magnetic impurities tends to weaken or destroy superconductivity. 
In 1960 Abrikosov and Gor'kov developed a theory of impurities 
in superconductors as pair breakers \cite{AG1960}, which treated impurity scattering 
in the Born approximation and was therefore limited to weak coupling. 
Subsequently, attempts to incorporate the Kondo effect have been made by many authors 
\cite{Soda01091967,Zittartz1970,MullerHartmann1971,Matsuura01031977,Matsuura01061977}. 
Shiba \cite{Shiba01091968} showed in an exactly solvable model of a classical impurity that there is a mid-gap 
excited bound state whose energy crosses zero as the interaction strength is varied. Closely related analyses 
were performed by Yu \cite{Yu1965} and Rusinov \cite{Rusinov1969}, and the mid-gap state 
is nowadays called the Yu-Shiba-Rusinov state, or simply, the Shiba state. 
Later, Sakurai \cite{Sakurai01121970} correctly interpreted the above result 
as a level crossing between the singlet state and the (unscreened) local moment state.  
A comprehensive NRG study \cite{Satori1992,Sakai1993} has finally 
confirmed that the predicted quantum phase transition 
indeed occurs at $T_{\rm K}/\Delta\sim 0.3$.  Recently the Shiba state and its transition 
have been observed in experiments using scanning tunneling microscopy \cite{Franke_Science_2011}. 
Realization of the Shiba state by ultracold fermionic superfluids is also proposed and actively investigated 
\cite{PhysRevA.83.033619,PhysRevA.83.061604,PhysRevA.83.063611}. 

The second setup to study the competition between Kondo screening and a depleted Fermi surface  is   
\emph{gapped} Fermi systems, such as insulators and semiconductors, 
where the density of states (DoS) at the Fermi level vanishes. 
Although gapped Fermi systems share similarity with 
superconductors, there are important differences as well (e.g., 
a classical spin in gapped Fermi systems shows no analogue of the 
transition (level crossing) induced by a classical spin in superconductors \cite{Ogura1993}.)  
So far, most of investigations were conducted on the gapped Anderson impurity model 
\cite{Saso1992,Takegahara1992,Ogura1993,ChenJayaprakash1998,
Galpin2008,GalpinLogan2008,MocaRoman2010,Pinto2012,Dasari2015} and 
it turned out that, in the presence of exact particle-hole symmetry, an arbitrarily small 
nonzero gap is enough to quench the Kondo effect at $T\to 0$. 
When the particle-hole symmetry is explicitly broken, the gap must overcome a nonzero threshold 
of order $T_{\rm K}$ to quench the Kondo effect.

The third class of systems where Kondo screening is suppressed is the so-called 
\emph{pseudogap} Fermi systems. Withoff and Fradkin \cite{PhysRevLett.64.1835} 
considered a magnetic impurity coupled to electrons with a DoS 
$N(E)\propto |E-E_F|^r$ ($r>0$) that vanishes at the Fermi level. 
This model was introduced to describe an impurity in $d$-wave superconductors, 
in which quasiparticles obey relativistic dispersions near the node of the gap. 
The upshot of \cite{PhysRevLett.64.1835} was that the Kondo effect 
occurs only when the impurity's coupling to the host exceeds a nonzero critical 
value that depends on $r$. Their conclusion was tested by subsequent studies 
\cite{PhysRevB.46.9274,Borkowski1994,ChenJayaprakash1995,PhysRevB.53.15079,
Ingersent1996,PhysRevB.56.11246,Bulla1997,Ingersent1998,FritzVojta2004,Glossop2005}, 
where the important role of particle-hole symmetry was pointed out.  
Dirty semiconductors with a Dirac cone were also investigated \cite{Fradkin1986a,Fradkin1986b}.  

In late years, a plethora of novel pseudogap Fermi systems have emerged: graphene \cite{Fritz2013graphene}, 
topological insulators (TI) with gapless surface states \cite{Hasan:2010xy,Qi:2011zya},  
and Dirac/Weyl semimetals \cite{Vafek:2013mpa,Wehling:2014cla}.%  
\footnote{The Dirac/Weyl semimetals with a linear dispersion in $d$ spatial 
dimensions correspond to a pseudogap system with the aforementioned exponent $r=d-1$.}   
Superconductivity in Dirac materials has become a subject of intensive theoretical 
research \cite{Kopnin2008PRL,Cho2012,Wei2014,Yang2014PRL,Bednik2015}. 
While the Kondo effect in (doped) Dirac/Weyl semimetals has already been studied  
\cite{Principi2015,Yanagisawa2015,MitchellFritz2015,Sun2015},    
the outcome of a gap opening at the Fermi energy due to Cooper pairing is not considered yet. 
Excited states induced by an impurity in the \emph{bulk} of TI were also studied \cite{LiuMa2009,Lu2012,Kuzmenko2014}, 
where exotic properties unseen in conventional insulators were reported.%
\footnote{In \cite{LiuMa2009,Lu2012,Kuzmenko2014} the excitation gap in the bulk 
has the form of a Dirac mass, rather than a Majorana mass generated from Cooper pairing.} 

Recently, the Kondo effect has attracted attention in the context of 
nuclear physics and Quantum Chromodynamics (QCD) at finite baryon density 
\cite{Yasui:2013xr,Hattori:2015hka,Yasui:2016ngy,Yasui:2016svc}. 
Needless to say, Dirac fermions appear naturally in high-energy physics because relativistic 
effects are important. In QCD, there is a long history of research on 
heavy flavors such as charm and bottom quarks with 
a large Dirac mass compared to the typical QCD scale 
\cite{Neubert:1993mb,Manohar2000Book,Hosaka:2016ypm}. 
In nuclear medium or quark matter, heavy quarks or 
hadrons behave as impurities and may experience the Kondo effect 
through interactions with the medium (in the color or isospin channel) 
\cite{Yasui:2013xr,Hattori:2015hka,Yasui:2016ngy,Yasui:2016svc}.  
What makes the impurity problem in QCD unique is the interplay of multiple 
quantum numbers (chirality, color, flavor, and spin). In QCD at high quark density, 
the Fermi surface is destabilized by Cooper instability mediated by attractive 
interactions between quarks and the ground state at low temperature is 
believed to be a color superconductor \cite{Rajagopal:2000wf,Alford:2007xm}. 
The presence of a BCS gap for quarks is detrimental to Kondo screening and 
understanding of their interplay is an intriguing nontrivial problem. However, 
color superconductivity has not been taken into account in the previous studies 
\cite{Yasui:2013xr,Hattori:2015hka,Yasui:2016ngy,Yasui:2016svc}.  

In this work,  we embark upon a study of competition 
between (color) superconductivity and Kondo screening in systems in which
the conduction fermions have relativistic dispersions.
Since various sources on channel  degrees of freedom
such as chirality, color, flavor, and spin
can be considered depending on setups, such systems in the absence of a superconducting gap
are prone to provoke the overscreened Kondo effect whose
realizations in condensed matter are few. In the presence of a superconducting gap,
a phase structure in the system is shown to be rich due to competitions among different energy scales.
In particular, we point out that the QCD Kondo effect is suppressed in the high-quark-density limit in which
a large superconducting gap shows up and kills a flow towards the Kondo regime.
At the same time, we show that a large superconducting gap rather leads to the presence of
the mid-gap state localized in the vicinity of an impurity, i.e.,
Shiba states, which may be new excitations in color superconductivity. 
Overall, the impurity physics of relativistic fermions at weak coupling is 
in close similarity to that of nonrelativistic fermions, provided the chemical potential is nonzero 
and a finite DoS is available. The relativistic nature of fermions becomes visible only when the 
coupling between the impurity and host is sufficiently strong or the chemical potential is small so that 
the Dirac point is close to the Fermi surface. 
In addition, the chirality and color of quarks that are missing in nonrelativistic fermions catalyze 
the overscreened Kondo effect.   

The outline of this paper is as follows. 
Section \ref{sc:os1} reviews possible Kondo effects in the absence of a superconducting gap and
presents a scaling analysis for several relativistic models.
Section \ref{sc:os2} discusses the phase structure in the presence of a gap based on
the scaling and existing NRG analyses. Comments on the QCD Kondo effect are also given.
Section \ref{sc:mids} investigates a quantitative phase diagram of a Dirac superconductor
by means of the slave-boson mean-field theory. By using the $T$-matrix method for 
a classical impurity, we derive the spectrum of Shiba states in a Dirac superconductor.
In Sec.~\ref{sc:cicsq} a similar analysis is performed for a color superconductor.
Section \ref{sc:cl} presents our conclusion and perspective.
In \autoref{sc:ap} the scaling functions of relativistic models are derived from 
Feynman diagrams up to two loops.

\section{\label{sc:os}Overscreened Kondo effect in Dirac systems \sc{sc:os}}
\subsection{\label{sc:os1}Scaling analysis in the normal state}

\begin{figure}[tb]
	\centering
	\includegraphics[height=.13\textwidth]{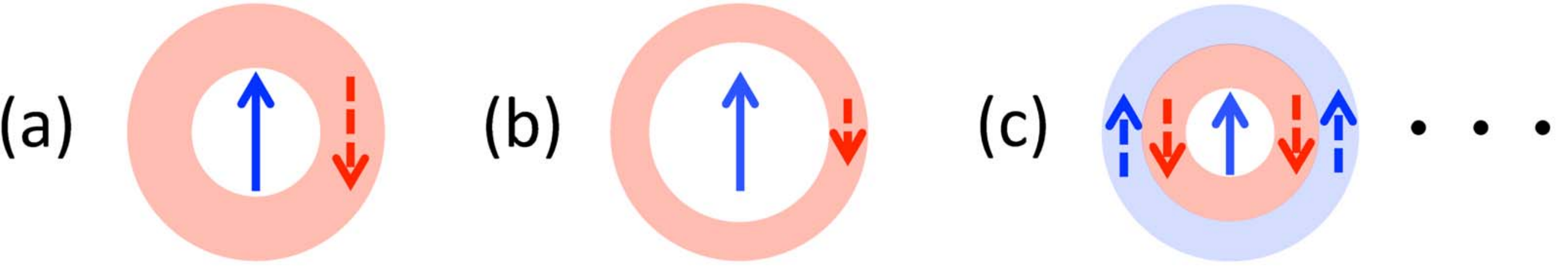} 
	\caption{\label{fg:Kscreen}Illustration of distinct Kondo effects:  
	(a) exact screening, (b) underscreening and (c) overscreening. 
	See the main text and Table~\ref{tb:phases} for further details.}
\end{figure}

Let us begin with a summary of known results. 
The most canonical setup of the Kondo effect consists of a spin-$1/2$ impurity and 
single-channel conducting electrons with spin $1/2$. When they are coupled via an antiferromagnetic 
interaction $g$, the impurity magnetic moment is 
\emph{exactly screened} at temperatures 
below the Kondo scale $T_{\rm K}$ in a way consistent with Landau's Fermi 
liquid theory. This phenomenon reflects the asymptotic freedom of renormalized 
coupling as was shown by Anderson with poor man's scaling \cite{Anderson1970};  
the flow of the effective coupling $g(D)$ as a function of the band width $D$ is 
governed by a scaling function $\beta(g)\equiv\dd g/\dd \log D \propto -g^2$. This negative 
$\beta$ function implies that, as $D$ is lowered, $g$ flows to the strong-coupling 
fixed point at infinity that corresponds to the phase with a vanishing impurity moment.  
The typical scale at which this crossover transition occurs is given by $T_{\rm K}$.  
In this \emph{exactly screened Kondo effect} [Fig.~\ref{fg:Kscreen}(a)], 
both the residual entropy and the residual magnetic moment at $T=0$ are zero. 

There are also cases in which the impurity spin is not exactly compensated 
\cite{Nozieres1980}. If the impurity spin is $S$ and the electrons have 
$M$ channels, we encounter qualitatively distinct phenomena depending on 
the relative magnitude of $S$ and $M/2$. When $S>M/2$, electrons are 
unable to screen the whole impurity moment, a situation called the \emph{underscreened 
Kondo effect} [Fig.~\ref{fg:Kscreen}(b)]. After a RG step, the residual magnetic moment interacts with electrons 
at the next scale ferromagnetically, which becomes irrelevant at low energy and the system flows toward the 
free fixed point. The entropy and the local moment remain nonzero in the IR. The underscreened Kondo effect was experimentally confirmed 
in \cite{Roch2009PRL}. 

%%%%%%%%%%%%%%%%%%%%%%%%%%%%
%%%%%%%%%%%%%%%%%%%%%%%%%%%%
%%%%%%%%%%%%%%%%%%%%%%%%%%%%
%%%%%%%%%%%%%%%%%%%%%%%%%%%%

By contrast, for $S<M/2$ there are more electrons than is 
necessary to screen the impurity spin, hence the impurity moment 
is \emph{overcompensated} [Fig.~\ref{fg:Kscreen}(c)]. This time the residual moment couples 
\emph{antiferromagnetically} to electrons at the next scale, rendering the free fixed point unstable. 
As was first noted by Nozi$\grave{\rm e}$res and Blandin \cite{Nozieres1980}, the IR 
limit in this case is governed by a nontrivial intermediate fixed point and exhibits a 
non-Fermi-liquid behavior.  A salient feature of overscreening is a vanishing 
magnetic moment and a \emph{nonzero} entropy in the IR 
\cite{Andrei1984,Affleck:1991tk,Gan1993PRL} (see Table~\ref{tb:phases}). 
A huge body of work has been developed for investigation of this \emph{overscreened Kondo effect}, 
as reviewed in \cite{Schlottmann1993review,CoxJarrell1996,CoxZawadowski1998}.  
Solutions of the multichannel Kondo and Anderson models have so far 
been obtained with the Bethe Ansatz \cite{Andrei1984,Tsvelick1984}, large-$N$ methods 
\cite{PhysRevLett.71.1613,Parcollet1997PRL,Parcollet1997PRB}, conformal  
field theory \cite{Affleck:1990by,*Affleck:1990iv,Affleck:1991tk,*Ludwig:1991tm},
and NRG~\cite{Bulla:2008zz}.  
The multichannel Coqblin-Schrieffer model was also solved with the Bethe Ansatz \cite{Jerez1998}. 
The multichannel \emph{pseudogap} models, which are of direct physical importance 
to $d$-wave superconductors and graphene, were also investigated in 
\cite{PhysRevB.56.11246,Ingersent1998,Vojta2001PRL,Schneider2011}. 
The two-channel Kondo effect was observed in a quantum-dot experiment 
\cite{Potok2007Nature}.  
 
Now we ask the main question in this section: 
in systems such as Dirac metals and quark matter that accommodate Dirac fermions, 
what kind of Kondo effect is caused by impurities? (For the moment, let us ignore the 
possibility of gap opening at the Fermi surface.) Recent studies 
\cite{Yasui:2013xr,Principi2015,Yanagisawa2015,MitchellFritz2015,
Sun2015,Hattori:2015hka,Yasui:2016svc} 
showed (i) that when the Fermi level coincides exactly with 
the Dirac point, the vanishing DoS suppresses any 
kind of Kondo effects unless the coupling exceeds a nonzero threshold, 
and (ii) that the Kondo effect sets in when the host fermions are doped. 

We would like to underline that it is actually the \emph{overscreened 
Kondo effect} that takes place in Case (ii). The following is a natural 
outcome of a physical reasoning based on the standard Kondo and Anderson 
model with nonrelativistic fermions: In Dirac metals, 
a single $1/2$ impurity would be screened exactly by a single 
Weyl fermion, and overscreened by a singlet Dirac fermion (equivalent to  
two degenerate Weyl cones). In quark matter with an impurity 
with $N_c$ colors, if the interaction only occurs through color, 
then a quark of one chirality would screen the impurity's color 
exactly; if the interaction occurs through both color and spin, 
then exact screening would be achieved by quarks with a single flavor, 
while quarks with more than one flavor would cause overscreening.  

To substantiate the above argument we employ two toy models.  
The first model is defined by the partition function  
$Z=\int{\cal D}[\psi,\psi^\dagger,\xi,\xi^\dagger] \exp\big(-\int \dd^4x~\mathcal{L}\big)$ 
with the Lagrangian
\ba
	\mathcal{L} 
%	& = \sum_{f=1}^{N_f}\psi^\dag_f (\der_\tau-\mu+i \bm{\sigma\cdot}\nabla)\psi_f  
%	+ \xi^\dag (\der_\tau-\mu_\xi)\xi  
%	+ \frac{G}{2}\sum_{f=1}^{N_f}\sum_{A=0}^{N_c^2-1}(\psi_f^\dag T^A \psi_f^{})(\xi^\dag T^A \xi) 
%	\\
	& = \sum_{f=1}^{N_f}\sum_{a=1}^{N_c}
	\psi^\dag_{fa} (\der_\tau-\mu+i \bm{\sigma\cdot}\nabla)\psi_{fa}  
	+ \sum_{a=1}^{N_c} \xi^\dag_a (\der_\tau-\mu_\xi)\xi_a 
	+ G \sum_{f=1}^{N_f} \sum_{a,b=1}^{N_c} 
	\psi_{fa}^\dag \psi^{}_{fb} \xi_{b}^\dag \xi_{a}^{} \,. 
	\label{eq:mdl}
\ea
This may be viewed as a relativistic analogue of the Coqblin-Schrieffer model \cite{PhysRev.185.847}. 
The fermionic impurity $\xi$ and the Weyl fermion $\psi$ (called quarks) 
are endowed with $N_c$ colors and the model has $\SU(N_c)$ internal symmetry. 
Both $\xi$ and $\psi$ transform in the fundamental representation of $\SU(N_c)$.  
In addition, $\psi$ has $N_f$ flavors and spin 1/2; $\xi$ has no spin.  
This is a crude model of a heavy quark immersed in a Fermi sea of light quarks 
in QCD. Analogous models were recently considered in 
\cite{Yasui:2013xr,Hattori:2015hka,Yasui:2016svc}.  
As regards quark matter with $u$ and $d$ quarks, the model \eqref{eq:mdl} 
with $N_f=2\times 2=4$ will apply. (The additional factor of $2$ accounts for the 
chirality of quarks.)   
The coupling $G>0$ mimics the color interaction mediated by gluons 
that is attractive in the color-antisymmetric channel. 
There is no spin-dependent interaction, reflecting the fact that spins of 
heavy quarks are frozen in QCD \cite{Isgur:1989vq,*Isgur:1989ed,Neubert:1993mb,Manohar2000Book}. 
$\mu>0$ and $\mu_\xi<0$ are chemical potentials 
for quarks and impurities, respectively. ($\mu>0$ is required to guarantee 
a nonzero DoS, a necessary condition for the Kondo effect.)
We adopt units in which $\hbar$, $k_{\rm B}$ 
and the Fermi velocity are all equal to unity. 

\begin{figure}[tb]
	\centering
	\includegraphics[width=.3\textwidth]{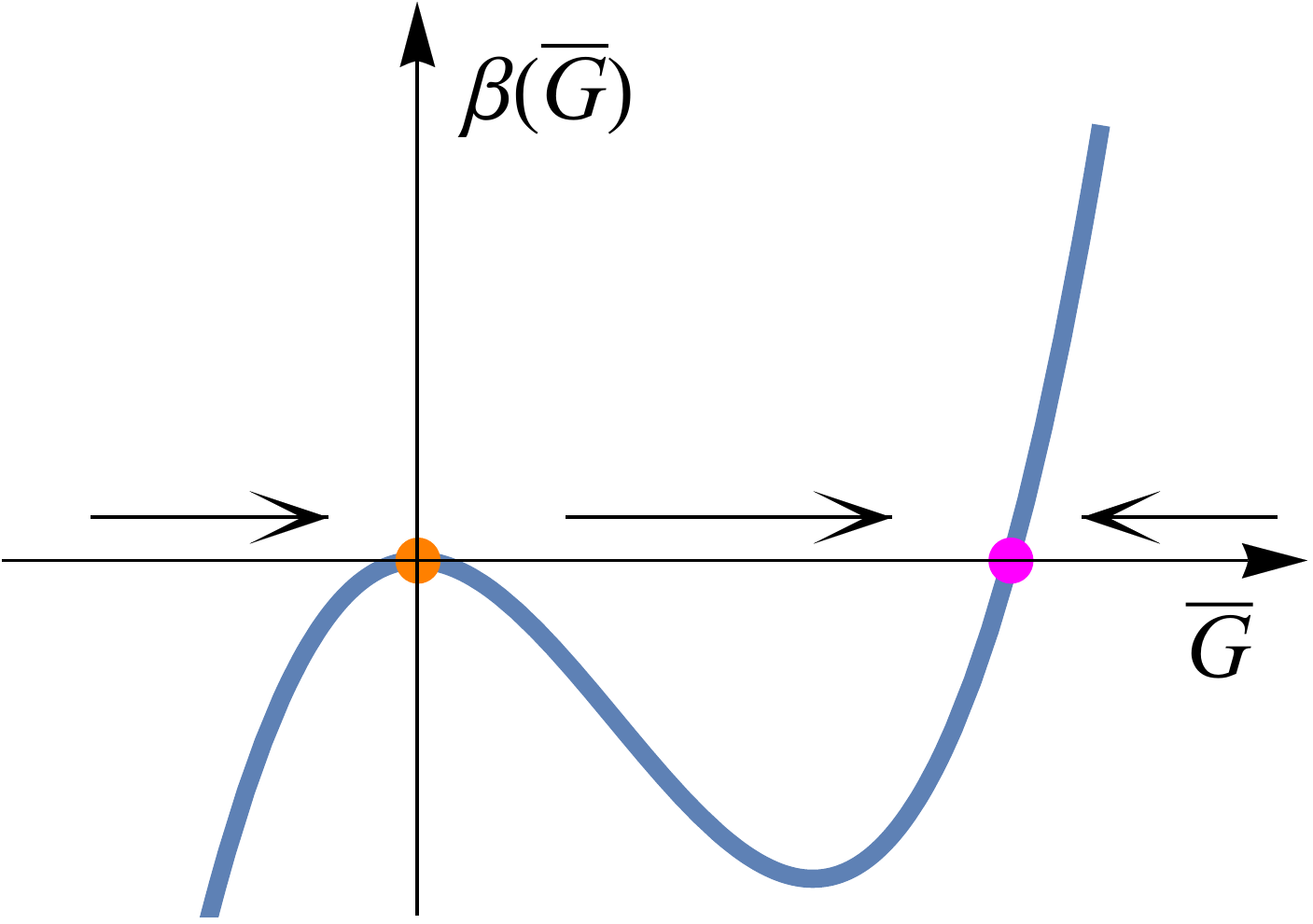}
	\vspace{-.5\baselineskip}
	\caption{\label{fg:betafunc}The scaling function of the dimensionless coupling 
	$\GG$. The arrows indicate the flow as $D\to 0$.}
\end{figure}

The scaling of an interaction towards strong coupling in the IR was originally 
shown by Anderson with the poor man's scaling \cite{Anderson1970}. 
Here we shall analyze the running of $G$ in the model \eqref{eq:mdl} for $N_f\gg 1$ 
with modern field-theoretical methods \cite{Peskin:1995ev}. 
As has been emphasized by Nozi$\grave{\rm e}$res and Blandin \cite{Nozieres1980}, 
it is mandatory to go to \emph{two loops} to expose the existence of an 
intermediate fixed point.  Namely, we determine 
the dependence of the renormalized dimensionless 
coupling $\GG\equiv  G \rho$ on the infrared cutoff $D$ at the two-loop level, 
where $D$ regularizes the singularity of the Fermi surface 
and $\rho=\mu^2/(2\pi^2)$ denotes the DoS at the Fermi surface. Our result is
\ba
	\beta(\GG) & \equiv \frac{\dd \GG}{\dd \log D}
	\\
	& = \frac{1}{2}N_c \GG^2(-1+N_f \GG)\,. 
	\label{eq:betaI}
\ea
The derivation is briefly sketched in \autoref{sc:ap}.%
\footnote{We refer the reader to \cite{HewsonBook,Gan1993PRL,Gan1994,Kuramoto1998,Bensimon2006,Aron2015} 
for full technical details of the scaling analysis at the next-to-leading order in the Kondo 
and Anderson models.} 
The beta function has a nontrivial fixed point $\GG=1/N_f$ 
besides the trivial fixed point $\GG=0$. In the limit $D\to 0$ the latter is 
apparently unstable, and the renormalization-group flow is attracted to 
the intermediate fixed point (Fig.~\ref{fg:betafunc}). 
%%%%%%%%%%%%%%%%%%%%%%%%%%%%
%%%%%%%%%%%%%%%%%%%%%%%%%%%%
%%%%%%%%%%%%%%%%%%%%%%%%%%%%
%%%%%%%%%%%%%%%%%%%%%%%%%%%%
For $N_f\gg 1$ the fixed point is at weak coupling and is inside the domain of 
validity of the perturbation theory. 
Therefore, for many flavors, the coupling is expected to flow from the weak 
limit at high $T$ to the intermediate fixed point at low $T$ where thermodynamic quantities obey 
scaling laws with nontrivial critical exponents. The characteristic 
scale for this crossover transition is given by the Kondo temperature, which 
is readily obtained as a renormalization-group invariant \cite{HewsonBook} 
\ba
	T_{\rm K} & \sim D \, [\GG(D)]^{2N_f/N_c} 
	\exp\Big(\mbox{$-\frac{2}{N_c \GG(D)}$}\Big)\,.
\ea
We next consider a simpler model where the Dirac fermions and impurities 
have no color, interacting only through spins just as in the conventional 
Coqblin-Schrieffer model. The Lagrangian is given by
\ba
	\mathcal{L} 
	& = \sum_{f=1}^{N_f}\psi^\dag_f (\der_\tau-\mu+i \bm{\sigma\cdot}\nabla)\psi_f  
	+ \sum_{s=1}^{2} \xi_s^\dag (\der_\tau-\mu_\xi)\xi_s + G \sum_{f=1}^{N_f} \sum_{s,s'=1}^{2} 
	\psi_{fs}^\dag \psi^{}_{fs'} \xi_{s'}^\dag \xi_{s}^{} 
	\label{eq:mdl2}
\ea
with $G>0$ and $\mu>0$. Both $\xi$ and $\psi_f$ have spin $1/2$.  
An analogous model was studied recently in \cite{Principi2015} 
(see also \cite{Yanagisawa2015,MitchellFritz2015,Sun2015} for related works). 
A three-dimensional Dirac metal with $K$ Dirac cones 
would correspond to the model \eqref{eq:mdl2} with $N_f=2K$. 
The scaling function for the coupling can be obtained in the same manner as for the previous model,
\ba
	\beta(\GG) & = \GG^2\mkakko{-1+\frac{N_f}{2} \GG} . 
	\label{eq:betaII}
\ea
The Kondo scale reads $T_{\rm K}\sim D\,\ee^{-1/\GG} \GG^{N_f/2}$.  Therefore 
again the beta function has a behavior depicted in Fig.~\ref{fg:betafunc} and the low-energy 
physics of the impurity at $N_f\gg 1$ is governed by a nontrivial fixed point. Note that this is true 
only for an antiferromagnetic coupling ($G>0$). The IR conformality in those two models offers an 
amusing prediction on the non-Fermi-liquid behavior of magnetic/colorful impurities 
in Dirac metals and quark matter and is worthy of investigation in future experiments.%
\footnote{The possibility of overscreening by Dirac fermions in graphene has already 
been discussed in the literature \cite{Fritz2013graphene}. We are unaware of 
a similar proposal in three dimensions.}%
\footnote{In atomic nuclei, the small size of the system acts as an IR cutoff and stops the flow to the 
fixed point, which would make the observation of overscreening difficult.}%
\footnote{Unlike quarks in high-energy physics, the dispersion relations of 
Dirac fermions in solid materials are not exactly linear \cite{Vafek:2013mpa,Wehling:2014cla}. 
Although the nonlinearity is neglected in the present study we suspect it will not cause 
a qualitative difference on low-energy physics in the bulk.} 
While we have so far considered interactions in either spin or color channels, 
the overscreened Kondo effect may well be also triggered by isospin-exchange interactions 
considered in \cite{Yasui:2013xr,Yasui:2016ngy}.   

There are three important caveats on the present argument. First, since the fixed point 
moves to strong coupling for smaller $N_f$, we are unable to precisely locate 
the boundary of the overscreened Kondo phase within perturbation theory. 
As an educated guess we conjecture that, in both models considered above, 
\emph{the overscreened intermediate fixed point would always exist for $N_f>1$, whilst  
the $N_f=1$ case would undergo exact screening characterized by a strong coupling 
to the host material in the IR.} This speculation is consistent with what is known for  
the $\SU(N)$ Kondo model \cite{PhysRevLett.71.1613,
Parcollet1997PRL,Parcollet1997PRB,Jerez1998,Bensimon2006} whose 
interaction Hamiltonian shares essentially the same structure as 
\eqref{eq:mdl} and \eqref{eq:mdl2}.  In the model \eqref{eq:mdl} 
with $N_f=1$, the ground state will be a color singlet 
$\epsilon^{a_1a_2\dots a_N}\xi_{a_1}\psi_{a_2} \psi_{a_3}\dots\psi_{a_{N}}$ 
formed by an impurity and $N-1$ quarks of different colors; 
if $\xi$ belongs to the anti-fundamental 
representation, the singlet $\xi_a\psi_a$ will form. More generally, 
the $N_f=1$ case would always lead to exact screening when $\xi$ belongs to an 
antisymmetric tensor representation of $\SU(N)$ \cite{Parcollet1997PRB}. 
By contrast, in all the $N_f>1$ cases we will encounter overscreening, provided the 
flavor symmetry is kept intact; the intermediate fixed point is weak against 
channel anisotropy \cite{Nozieres1980}. In dense QCD, however, the 
$\SU(N_f)$ symmetry of light quarks is slightly violated by current quark masses and 
electromagnetic interactions, which may ultimately result in a suppression of 
the overscreened Kondo effect in the far-IR limit. This point deserves further study.%
\footnote{The effect of isospin-symmetry breaking on the Kondo effect of nucleons 
was discussed in Sec.~IV.A of \cite{Yasui:2016ngy}.} 

Secondly, in this work we neglect correlations between localized moments. 
In strongly correlated electron systems, the so-called Ruderman-Kittel-Kasuya-Yosida (RKKY) 
interaction \cite{Ruderman1954,*Kasuya1956,*Yoshida1957} 
between localized moments induces a variety of exotic phases and its importance 
in Dirac/Weyl systems has been recently explored (see e.g., \cite{Araki2016RKKY}), but this 
goes beyond the scope of this paper.

As the third and last caveat, it should be remarked that the models \eqref{eq:mdl} 
and \eqref{eq:mdl2} treat Dirac fermions as non-interacting. This is a poor approximation 
to QCD at low and intermediate density, where non-Abelian gauge interactions among 
quarks are so strong that quarks acquire a large Dirac mass dynamically and are permanently 
confined inside color-singlet nucleons. Neither effects are incorporated into \eqref{eq:mdl}.  
At best the model \eqref{eq:mdl} would be a sensible analogy to quark matter 
only for the high-density region $\mu\gg \Lambda_{\rm QCD}$, but nevertheless the 
gauge interaction between quarks at the Fermi surface will inevitably trigger color 
superconductivity \cite{Bailin:1983bm,Rajagopal:2000wf,Alford:2007xm} 
and brings about numerous changes into the above consideration of impurity dynamics. 
We will analyze the competition between Kondo screening and 
Cooper pairing in Dirac systems 
later in Secs.~\ref{sc:mids} and \ref{sc:cicsq}.  Before delving into a quantitative analysis, 
we wish to briefly summarize our main conclusion in the next subsection.

\subsection{\label{sc:os2}Taxonomy of impurity states in multichannel (color) superconductors}

The Kondo effect rests on the availability of a nonzero DoS at the Fermi surface 
while both the (color-)superconducting gap and temperature spoil the Fermi 
surface. Naturally a competition among them arises and it has a long history 
of research as reviewed in Sec.~\ref{sc:intro}. It seems that most of the research 
conducted so far has focused on the case of \emph{exact screening}, leaving the 
domain of overscreened Kondo systems much less explored (but see 
\cite{Zitko2016} for a recent attempt to fill the gap). To set a stage for 
later discussions, we begin with a summary of phases in Kondo systems 
(Table~\ref{tb:phases}).  
The emergence of such rich phases as in Table~\ref{tb:phases} 
has been revealed in studies of the Kondo effect in multichannel 
gapped/pseudogap Fermi systems 
\cite{PhysRevB.56.11246,Ingersent1998,Vojta2001PRL,Schneider2011}.%
\footnote{The ASC is called LM$'$ in \cite{Schneider2011}.}   
%%%%%%%%%%%%%%%%%%%%%%%%%
%%%%%%%%%%%%%%%%%%%%%%%%%
\begin{table}[h]
	\centering
	\begin{tabular}{lcccccc}
		\hline \hline 
		&~~US~~ & ~~K~~ & ~~OS-FP~~ & ~~LM~~ & ~~ASC~~ & ~~SC~~
		\\\hline 
		Residual $T=0$ entropy ~~
		& $>0$ & $0$ & $>0$ & $>0$ & $>0$ & $>0$
		\\ 
		Residual impurity spin ~~
		& $>0$ & $0$ & $0$ & $>0$ & $0$ & $>0$ 
		\\\hline \hline 
	\end{tabular}
	\caption{\label{tb:phases}
	Classification of phases in the multichannel Kondo model. 
	(US:~underscreened phase, K:~normal Kondo phase with exact screening, 
	OS-FP:~overscreened fixed point, 
	LM:~local moment phase, ASC:~asymmetric strong-coupling phase, 
	SC:~strong-coupling limit.) ASC and SC reduce to K in a single-channel model.}
\end{table}
%%%%%%%%%%%%%%%%%%%%%%%%%
%%%%%%%%%%%%%%%%%%%%%%%%%

Here the local moment phase denoted by ``LM'' is a free spin state 
with no Kondo screening. The asymmetric strong coupling phase labeled as ``ASC'', 
which appears only when the particle-hole asymmetry is sufficiently strong \cite{Ingersent1998}, 
is a phase where the impurity spin is completely screened by a minimal number of 
electrons; for instance, in the $K$-channel Kondo model with a spin-$1/2$ impurity, ASC is 
a phase where only one of the $K$ electrons participates in the screening and the other $K-1$ are decoupled, 
resulting in a $K$-fold degenerate ground state and a residual entropy $S=\log K$. 
This ASC phase, first identified in \cite{Ingersent1998} with NRG, is 
absent in metallic Kondo and Anderson models but can exist stably in pseudogap models 
if the coupling is sufficiently strong.   
The ``SC'' is a phase where fermions in all channels couple 
to the impurity symmetrically. This corresponds to the strong-coupling limit of 
the $K$-channel $\SU(N)$ Kondo model with an impurity 
in the fundamental representation of $\SU(N)$: at the infinite-coupling fixed point  
of this model, $(N-1)K$ electrons combine with the impurity spin to form a larger 
representation of $\SU(N)$ \cite{Parcollet1997PRB,Jerez1998}. In models with a 
metallic DoS, this fixed point is unstable and the RG flow is attracted to the overscreened 
intermediate fixed point \cite{Nozieres1980}; however, in a gapped model at sufficiently 
strong coupling this SC phase can be stable, as was numerically shown in \cite{Zitko2016}.  
The other phases (US, K and OS-FP) were described in the previous subsection.  
As a quick guide, below we depict phases in a two-channel model,  
denoting the impurity spin $1/2$ by $\Uparrow$ and the spin of electrons by $\uparrow$: 
\ba
	|\text{LM}\rangle=|\!\Uparrow\rangle\,, \quad
	|\text{ASC}\rangle=|\!\Uparrow\downarrow\rangle\,, \quad 
	|\text{SC}\rangle=|\!\Uparrow\downarrow\downarrow\rangle\,. 
	\label{eq:iroiro}
\ea

\begin{figure}[tb]
	\centering
%	\includegraphics[height=.22\textwidth]{fg_pd_A} \qquad \quad \quad 
%	\includegraphics[height=.22\textwidth]{fg_pd_B} \qquad \quad \quad  
%	\includegraphics[height=.22\textwidth]{fg_pd_C}
%	\put(-465,95){\large $(a)$}
%	\put(-303,95){\large $(b)$}
%	\put(-142,95){\large $(c)$}
	\includegraphics[height=.23\textwidth]{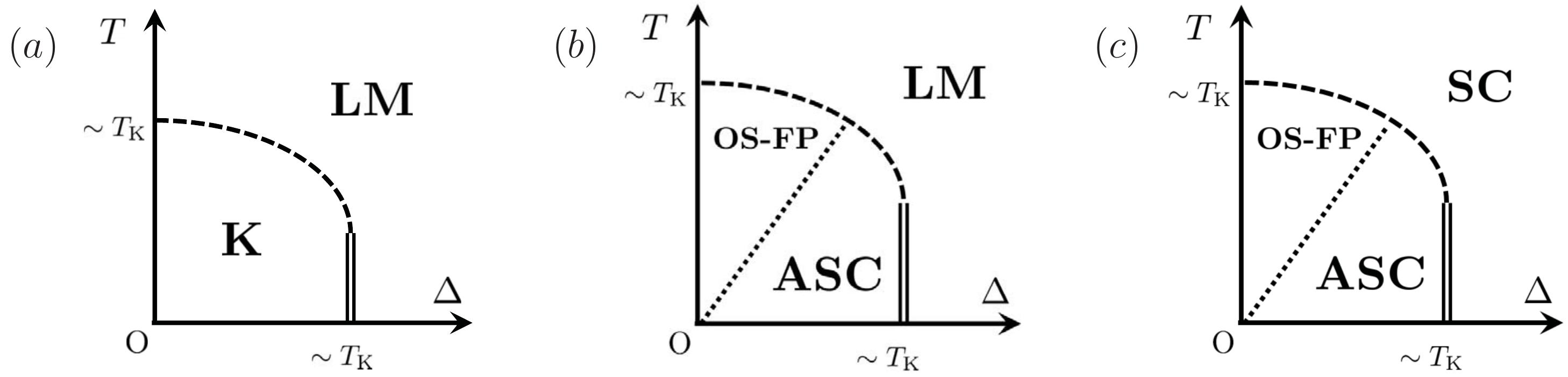}
	\vspace{-.5\baselineskip}
	\caption{\label{fg:OSp}Schematic phase diagram of Kondo impurities in 
	Dirac superconductors for (a) $N_f=1$, (b) $N_f=2$ at weak coupling, and 
	(c) $N_f=2$ at strong coupling. The double line ($=\!=\!=$) represents a first-order transition.}
\end{figure}

Concerning the $N_f=1$ case of the Dirac models \eqref{eq:mdl} and \eqref{eq:mdl2}, 
it is not difficult to estimate physical effects of a superconducting gap $\Delta$ 
for the host fermions $\psi$ on Kondo screening. On the basis of knowledge on 
Kondo impurities in gapped \cite{Saso1992,Takegahara1992,Ogura1993,ChenJayaprakash1998,
Galpin2008,GalpinLogan2008,MocaRoman2010,Pinto2012,Dasari2015} and 
pseudogap \cite{PhysRevLett.64.1835,PhysRevB.46.9274,Borkowski1994,
ChenJayaprakash1995,PhysRevB.53.15079,Ingersent1996,PhysRevB.56.11246,
Bulla1997,Ingersent1998,FritzVojta2004,Glossop2005} Fermi systems as well as  in 
$s$-wave superconducting hosts \cite{Shiba01091968,Sakurai01121970,Satori1992,Sakai1993}, 
one can draw a phase diagram for Kondo impurities in the $(T,\Delta)$-plane, 
as shown in Fig.~\ref{fg:OSp}(a). Here $\Delta$ is treated as an externally tunable parameter. 
Since $T_{\rm K}$ is the only dynamical scale in this problem, the phase boundary should 
be roughly set by $T_{\rm K}$ up to $\calO(1)$ numerical factors that depend on model details.%
\footnote{The zero-temperature transition occurs at $\Delta\simeq 3.3T_{\rm K}$ in the spin-1/2 
single-channel Kondo model \cite{Satori1992}.} At $T=0$ a first-order transition occurs due to 
level crossing between a screened state and an unscreened state \cite{Sakurai01121970,Satori1992}. 
This phase diagram will be revisited with a mean-field theory in Sec.~\ref{sc:mids}. 

%%%%%%%%%%%%%%%%%%%%%%%%%%%%
%%%%%%%%%%%%%%%%%%%%%%%%%%%%
%%%%%%%%%%%%%%%%%%%%%%%%%%%%
%%%%%%%%%%%%%%%%%%%%%%%%%%%%

Next we turn to a more nontrivial question: 
what occurs in a multi-flavor system in the presence of a pairing gap? 
If $\Delta=0$, physics at low $T$ is of course described by OS-FP. On the other hand, 
at $T=0$ the intermediate fixed point tends to be washed out as soon as $\Delta\ne0$ is 
turned on, giving place to ASC \cite{Ingersent1998,Vojta2001PRL,Schneider2011}. Therefore one expects 
OS-FP and ASC to dominate the phase diagram at low $T$ and small $\Delta$. 
When the condition $\Delta\ll T < T_{\rm K}$ is met, the overscreening will 
not be affected by the tiny gap. As $\Delta$ grows, OS-FP would gradually be 
preempted by ASC. Assuming, as a simplest scenario, that the boundary between the two 
regimes extends along a straight line $T\propto\Delta$, we arrive at the phase diagram  
in Fig.~\ref{fg:OSp}(b). This phase diagram, valid at weak coupling, is consistent with 
a recent NRG study of the two-channel Kondo model with a BCS gap \cite{Zitko2016},  
in which the noncommutativity of limits $T\to 0$ and $\Delta\to 0$ was clearly observed. 
To map out the phase diagram at strong coupling, we recall that the two-channel Kondo model 
has a strong-weak duality \cite{KolfKroha2007}. In this mapping, LM and SC are interchanged. 
This duality is implied by the fact that both LM and SC have spin $1/2$ [cf.~\eqref{eq:iroiro}]. 
In Fig.~\ref{fg:OSp}(c) we present a phase diagram at strong coupling, obtained from 
Fig.~\ref{fg:OSp}(b) by interchanging LM and SC. Notably, at $T=0$ the quantum phase 
transition occurs between ASC and SC (instead of ASC and LM), which has been confirmed 
in a NRG study \cite{Zitko2016}.  Since the duality above is specific to the two-channel 
problem, the task of mapping out a phase diagram for $N_f>2$ is far more challenging. 
In that case, one has to take into account new states that are intermediate between 
$|\text{ASC}\rangle=|\!\Uparrow\downarrow\rangle$ and 
$|\text{SC}\rangle=|\!\Uparrow\underbrace{\downarrow\dots\downarrow}_{N_f}\rangle$, 
which may possibly lead to multiple quantum phase transitions. 
 
Finally we wish to comment on the QCD Kondo effect \cite{Hattori:2015hka}. 
In the high-density limit the renormalized gauge coupling $g$ is small and a weak-coupling computation 
can be used to estimate the leading behavior of observables \cite{Collins:1974ky}. 
It was demonstrated in \cite{Hattori:2015hka} via a one-loop RG analysis 
for normal-state quark matter that the Kondo effect for colors of a heavy impurity 
does take place through a color-exchange interaction at the Kondo scale%
\footnote{The Kondo scale in \cite{Hattori:2015hka} is defined as the energy 
scale at which the one-loop scattering amplitude between light quarks and a heavy quark diverges. 
Note that, while the Kondo temperature in solids is usually a monotonically increasing 
function of the DoS at the Fermi energy, the dependence of $\Lambda_{\rm K}$ 
on $\mu$ is nontrivial due to the running of the renormalized coupling $g$.} 
\ba
	\Lambda_{\rm K} \propto \mu \exp\mkakko{ - \frac{8\pi^2}{N_cg^2} }.  
	\label{eq:LamK}
\ea
Their analysis neglected interactions between light quarks. In reality, 
gluons mediate attractive interactions between quarks in a color-antisymmetric channel 
and inevitably induce color superconductivity 
with a BCS gap $\Delta$ for quarks. The dependence of $\Delta$ on $g$ 
has been computed as $\Delta\sim \mu g^{-5} \ee^{-c/g}$ with $c=3\pi^2/\sqrt{2}$ for 
$N_c=3$ and $c=2\pi^2$ for $N_c=2$   
\cite{Son:1998uk,Hong:1999fh,Schafer:1999jg,Pisarski:1999bf,Pisarski:1999tv}. 
The hierarchy of scales $\Delta\gg \Lambda_{\rm K}$ 
at $g\ll 1$ indicates that the QCD Kondo effect is suppressed by quark pairing 
in quark matter at asymptotically high density.%
\footnote{Here the impurity is assumed to be infinitely heavy. In reality, 
heavy flavors ($c,b$) carry a finite mass of order $1\sim 4$~GeV and 
they will be populated at asymptotically high baryon density. This physical 
limit is outside the consideration here.} 
If we start from the quark-gluon plasma phase 
at high $T$ and goes down in temperature, quarks will acquire a gap 
at $T\sim \Delta$ and hence the flow of the coupling is cut off far before 
the Kondo regime $T\lesssim T_{\rm K}$ is reached. 

A few supplementary remarks are in order. 
\begin{enumerate}
	\item 
	The above argument alone does not preclude the 
	QCD Kondo effect at intermediate quark density. We will take up 
	this issue in Sec.~\ref{sc:cicsq} in a mean-field theory. 
	\item 
	Even when the Kondo screening is quenched, it leaves behind a fingerprint:  
	the coupling of an impurity to host quarks gives rise to localized 
	excited states, the so-called \emph{Shiba states}, 
	that lie inside the spectral gap of quasiparticles \cite{RevModPhys.78.373}. 
	Detailed analyses will be given in Secs.~\ref{sc:mids} and \ref{sc:cicsq} below. 
	\item 
	The suppression of Kondo screening will be robust in phases of quark matter 
	where all quark species acquire a gap (Majorana mass), such as the color-flavor-locked (CFL) phase  
	of three-flavor QCD \cite{Alford:1998mk}, the high-isospin-density limit of two-flavor QCD 
	\cite{Son:2000xc,Son:2000by} and the color-spin-locked phase of one-flavor QCD \cite{Schafer:2000tw,Schmitt:2004et}; 
	in all these phases the gap parametrically depends on $g$ as $\Delta\sim \mu g^{-5} \ee^{-c/g}$. 
	By contrast, the two-flavor color-superconducting phase (2SC) is more subtle because 
	quarks with one out of three colors remain gapless \cite{Alford:1997zt,Rapp:1997zu}.    
	Since the gapless quarks are neutral under the residual unbroken gauge group $\SU(2)\subset \SU(3)$, 
	their interaction with gapped quarks and impurities is suppressed at low energy and 
	will not modify our conclusion that the Kondo effect is eliminated by the pairing gap. 
	\item 
	One may wonder if the Kondo effect based on a color-exchange interaction is 
	actually well-defined in the CFL phase, where the color gauge group $\SU(3)_{\rm C}$ is 
	completely broken by diquark condensates via the Anderson-Higgs mechanism. Note however that 
	the diagonal subgroup $\SU(3)_{\rm C+L+R}$ of color and flavor symmetries  
	$\SU(3)_{\rm C}\times\SU(3)_{\rm L}\times\SU(3)_{\rm R}$ is unbroken in the CFL phase 
	(as long as the strange quark mass can be neglected, of course). Quarks then transform 
	in the singlet and octet representation of $\SU(3)_{\rm C+L+R}$, whereas a heavy 
	quark belongs to the fundamental representation. Since host fermions transform in a larger 
	representation of the symmetry group than the impurity, the \emph{overscreened} Kondo effect 
	will take place, provided the pairing gap is small. (A similar situation for $\SU(2)$ spin was 
	studied in \cite{FabrizioZarand1996,SenguptaKim1996}.) The Kondo effect we claim here to be suppressed 
	should be interpreted as this overscreened Kondo effect in $\SU(3)_{\rm C+L+R}$\,.
	\item 
	We have so far argued that the color moment of heavy quarks will not be Kondo-screened 
	in a fully gapped phase of dense quark matter. This is not a full story if a non-Abelian 
	subgroup of a gauge group is left unbroken after quark pairing: the color degrees must be permanently 
	confined into color-singlet hadrons \cite{Greensite2011Book}. This issue arises in the 
	2SC phase of three-color QCD, the high-isospin-density phase of two-flavor QCD with any colors, 
	and the superfluid phase of two-color QCD. 
	Although both the Kondo effect and color confinement concern the screening 
	of impurity's color moment, they are intrinsically different mechanisms: Kondo screening is 
	a Fermi surface effect, whereas quark confinement originates from strong-coupling dynamics of 
	gluons and has no bearing on the Fermi surface. What happens to an impurity's color moment 
	if it is not subject to Kondo screening? As was shown in \cite{Rischke:2000cn}, 
	the energy scale of confinement in the 2SC phase $\Lambda'_{\rm QCD}$ is significantly 
	lowered from the strong-coupling scale in the QCD vacuum due to medium effects. Parametrically  
	$\Lambda'_{\rm QCD}\sim \Delta \exp[-\text{const.}\times \mu/(g\Delta)]$, which is 
	even smaller than the Kondo scale $\Lambda_{\rm K}$ in \eqref{eq:LamK}.  
	This implies a separation of scales $\Lambda'_{\rm QCD}\ll \Lambda_{\rm K}\ll \Delta$ at $g\ll 1$. 
	Therefore the impurity moment will be effectively free for the energy scale $E\gg\Lambda'_{\rm QCD}$ 
	but screened via the confinement mechanism for $E\lesssim \Lambda'_{\rm QCD}$; 
	the LM phase found in Fig.~\ref{fg:OSp} and our model analysis in Sec.~\ref{sc:cicsq} below 
	should be considered as valid only at low, but not too low, energy scales. 
\end{enumerate}

\section{\label{sc:mids}Impurity in Dirac superconductors \sc{sc:mids}}
\subsection{\label{sc:mf}Mean-field theory \sc{sc:mf}}
\subsubsection{\label{sc:mf1}Model setup \sc{sc:mf1}}

In this section we investigate the competition between superconductivity and 
Kondo screening in a model of three-dimensional Dirac fermions.
We use the slave-boson mean-field theory for the Kondo problem, developed in 
\cite{Read-Newns1983,Coleman1984} and reviewed in \cite{RevModPhys.59.845,Coleman:2015uma}.  
Early applications of the large-$N$ slave-boson technique to 
the Kondo problem in superconductors can be found in \cite{PhysRevB.46.9274,Borkowski1994}.  
More recently, the method was applied to the description 
of the Kondo effect in quark matter \cite{Yasui:2016svc}, albeit neglecting 
the effect of color superconductivity. 

In this section we employ the model \eqref{eq:mdl2} with $N_f=1$, i.e., just a single Weyl cone. 
In this benign case we need not incorporate the overscreened Kondo effect and 
the analysis is greatly simplified. Incorporating a Majorana mass $\Delta$ that represents the $s$-wave BCS gap, we have the action
(in natural units $\hbar=k_{\rm B}=1$) 
\ba
	S & = \int \dd\tau\,\dd^3x \Big[
	\psi^\dagger (\der_{\tau} - \mu + i v \bm{\sigma\cdot}\nabla) \psi 
	+ \frac{\Delta}{2} (\psi^\text{T}\sigma^2 \psi + 
	\psi^{\dagger}\sigma^2 \psi^*) 
	+ \xi_s^\dagger(\der_\tau - \mu_\xi) \xi_s 
%	+ \frac{G}{2} \sum_{\nu=0}^{3}
%	(\psi^\dagger \sigma^\nu \psi)(\xi^\dagger \sigma^\nu \xi) 
	+ G \psi^\dagger_s \psi^{}_{s'} \xi_{s'}^\dagger \xi_s
	\Big] \,,
	\label{eq:lagw0}
\ea
where $v$ is the Fermi velocity and $G>0$ as in Sec.~\ref{sc:os}. 
Next we perform the Hubbard-Stratonovich transformation
\ba
	- G (\psi^\dagger_s \xi_s) (\xi^\dagger_{s'} \psi_{s'}) 
	& ~\Rightarrow~ \frac{|V|^2}{G} + V^* (\psi_s^\dagger \xi_s) + V (\xi^\dagger_s \psi_s) 
\ea
with an auxiliary complex scalar field $V$.  In the mean-field approximation 
$V=\text{const.}$, one can rotate the global $\U(1)$ phase of $\xi$ so that $V\geq0$ 
without loss of generality. Then  
\ba
	S & = \int \dd^4x \bigg[
		\psi^\dagger (\der_{\tau} - \mu + i v \bm{\sigma\cdot}\nabla) \psi + 
		\frac{\Delta}{2} (\psi^\text{T}\sigma^2 \psi + 
		\psi^{\dagger}\sigma^2 \psi^*)
		+ \xi^\dagger (\der_\tau - \mu_\xi) \xi 
		+ V (\psi^\dagger \xi + \xi^\dagger \psi ) + \frac{V^2}{G}  
	\bigg]
	\\
	& = \int \dd^4 x \bigg[ \frac{1}{2}
	\bep \psi^\text{T} & \psi^\dagger & \xi^\text{T} & \xi^\dagger \eep
	\bep
		\Delta \sigma^2 & 
		\der_\tau+\mu+iv \bm{\sigma}^{\rm T}\!\bm{\cdot}\nabla  & 0 & - V \1_2
		\\
		\der_\tau-\mu+iv \bm{\sigma\cdot}\nabla & \Delta \sigma^2 & V \1_2 & 0
		\\
		0 & - V \1_2 &0& (\der_\tau + \mu_\xi) \1_2
		\\
		V \1_2 & 0 & (\der_\tau - \mu_\xi) \1_2 &0
	\eep
	\bep \psi \\ \psi^* \\ \xi \\ \xi^* \eep
	+ \frac{V^2}{G} \bigg]\,. \!\!
\ea
The thermodynamic potential per unit volume is obtained [with the Matsubara frequency $\omega_n\equiv (2n+1)\pi T$] as
\ba
	\Xi & = \frac{V^2}{G} - \frac{1}{2} T \sum_{n\in\ZZ} \int\frac{\dd^3k}{(2\pi)^3}
	\log \det \bep
		\Delta \sigma^2 & 
		i\omega_n+\mu- v \bm{\sigma}^{\rm T}\!\bm{\cdot k}  
		& 0 & - V \1_2
		\\
		i\omega_n - \mu - v \bm{\sigma\cdot k} & 
		\Delta \sigma^2 & V \1_2 & 0
		\\
		0 & - V \1_2 &0& (i\omega_n + \mu_\xi) \1_2
		\\
		V \1_2 & 0 & (i\omega_n - \mu_\xi) \1_2 &0
	\eep
	\\
	& = \frac{V^2}{G} - \frac{1}{2} T \sum_{n\in\ZZ} \int\frac{\dd^3k}{(2\pi)^3}
	\log \Big\{ [\omega_n^2+B_1^{+}(k)][\omega_n^2+B_1^{-}(k)]
	[\omega_n^2+B_2^{+}(k)][\omega_n^2+B_2^{-}(k)] 
	\Big\}\,, 
	\label{eq:thepote}
\ea
where $(k\equiv |\bm{k}|)$ 
\begin{subequations}
	\label{eq:BB}
	\ba
		B_1^{\pm}(k) 
		& = \frac{1}{2}\Big[
			(vk-\mu)^2+\Delta^2+2V^2+\mu_\xi^2 \pm 
			\sqrt{
				\big\{ (vk-\mu)^2+\Delta^2-\mu_\xi^2 \big\}^2 + 
				4V^2 \big\{ (vk-\mu-\mu_\xi)^2+\Delta^2 \big\}
			}\ 
		\Big] \,,
		\\
		B_2^{\pm}(k) 
		& = \frac{1}{2}\Big[
			(vk+\mu)^2+\Delta^2+2V^2+\mu_\xi^2 \pm 
			\sqrt{
				\big\{ (vk+\mu)^2+\Delta^2-\mu_\xi^2 \big\}^2 + 
				4V^2 \big\{ (vk+\mu+\mu_\xi)^2+\Delta^2 \big\}
			}\ 
		\Big] \,.
	\ea
\end{subequations}
The Matsubara sum in \eqref{eq:thepote} can be done with the formula 
\ba
	\sum_{n\in\ZZ} \log(\omega_n^2+B)\bigg|_{\omega_n=(2n+1)\pi T}
	= 2 \log \cosh \frac{\sqrt{B}}{2T}+\dots 
\ea
for $B>0$, where the dots represent divergent terms 
that are $T$-dependent but $B$-independent. This yields 
\ba
	\Xi & = \frac{V^2}{G} - T \int\frac{\dd^3k}{(2\pi)^3}
	\Bigg[
		\log\cosh \frac{\sqrt{B_1^{+}(k)}}{2T} + 
		\log\cosh \frac{\sqrt{B_1^{-}(k)}}{2T} + 
		\log\cosh \frac{\sqrt{B_2^{+}(k)}}{2T} + 
		\log\cosh \frac{\sqrt{B_2^{-}(k)}}{2T}
	\Bigg]
	\\
	& = \frac{V^2}{G} - \int_0^{\Lambda}\frac{\dd k\,k^2}{4\pi^2}
	\Big[
		\sqrt{B_1^{+}(k)} + \sqrt{B_1^{-}(k)} + 
		\sqrt{B_2^{+}(k)} + \sqrt{B_2^{-}(k)} + 
		2T \log \mkakko{ 1+\ee^{-\beta \sqrt{B_1^{+}(k)}} }
	\notag
	\\
	& \qquad 
		+ 2T \log \mkakko{ 1+\ee^{-\beta \sqrt{B_1^{-}(k)}} }
		+ 2T \log \mkakko{ 1+\ee^{-\beta \sqrt{B_2^{+}(k)}} }
		+ 2T \log \mkakko{ 1+\ee^{-\beta \sqrt{B_2^{-}(k)}} }
	\Big] \,,
	\label{eq:poten}
\ea
where $\Lambda$ is a UV cutoff (with $v\Lambda>\mu$ so that the Fermi surface is included) 
and we dropped irrelevant constants from the RHS.  In solid state physics,
the magnitude of $\Lambda$ is essentially set by the band width of conduction electrons. 
The values of $V$ and $\mu_\xi$ are determined by solving the constraints
\ba
	\label{eq:Xiextrema}
	\frac{\der}{\der V} \Xi = 0 \qquad 
	\text{and}\qquad \frac{\der}{\der \mu_\xi} \Xi = n_\xi \,.
\ea
In natural units, all dimensionful quantities can be measured with 
$v$ and $\Lambda$. It is convenient to attach a hat $\hat{~}$ to 
each dimensionless quantity as follows.
\ba
	\hat{\mu}=\frac{\mu}{v\Lambda}, \quad 
	\hat{\mu}_\xi=\frac{\mu_\xi}{v\Lambda}, \quad 
	\hat{T}=\frac{T}{v\Lambda}, \quad 
	\hat{\Delta}=\frac{\Delta}{v\Lambda}, \quad 
	\hat{V}=\frac{V}{v\Lambda}, \quad 
	\hat{G}=\frac{G\Lambda^2}{v}, \quad
	\hat{\Xi}=\frac{\Xi}{v\Lambda^4}, \quad 
	\hat{k}=\frac{k}{\Lambda}, \quad 
	\hat{B}_{1,2}^{\pm}=\frac{B_{1,2}^{\pm}}{(v\Lambda)^2}\,.
	\label{eq:hatnotation}
\ea

\subsubsection{\label{sc:mf2}Quasiparticle spectra \sc{sc:mf2}}

%%%%%%%%%%%%%%%%%%%%%%%%%%%%
%%%%%%%%%%%%%%%%%%%%%%%%%%%%
\begin{figure*}[bt]
	\centering
%	\includegraphics[width=.24\textwidth]{fg_disp_A} \qquad \quad
%	\includegraphics[width=.24\textwidth]{fg_disp_B} \qquad \quad
%	\includegraphics[width=.24\textwidth]{fg_disp_C} 
%	\put(-443,112){\large $(a)$}
%	\put(-288,112){\large $(b)$}
%	\put(-137,112){\large $(c)$}
	\includegraphics[height=.25\textwidth]{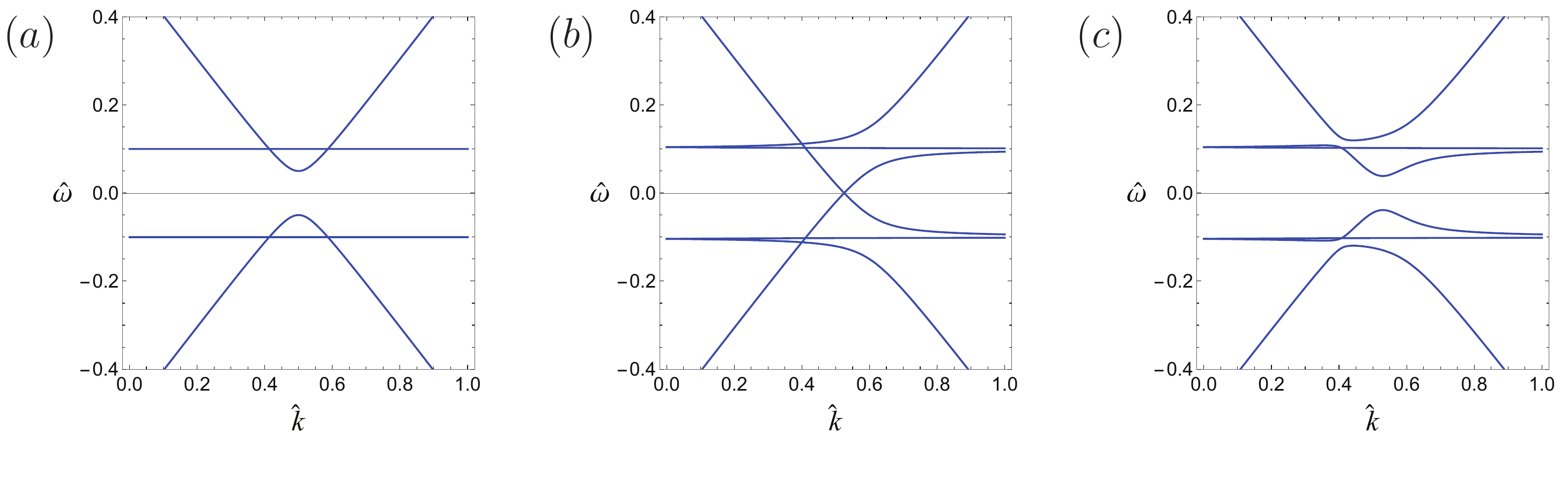}
	\vspace{-\baselineskip}
	\caption{\label{fg:disp}Dispersion relations of quasiparticles 
	$\big\{\hat\omega = \pm \sqrt{\hat B_1^{\pm}(\hat k)},\,\pm \sqrt{\hat B_2^{\pm}(\hat k)}\,\big\}$  
	for $\hat\mu=0.5$ and $\hat\mu_\xi=-0.1$.  
	(a):~$(\hat\Delta,\hat{V})=(0.05, 0)$, (b):~$(\hat\Delta,\hat{V})=(0, 0.05)$, 
	(c):~$(\hat\Delta,\hat{V})=(0.05, 0.05)$. }
\end{figure*}
%%%%%%%%%%%%%%%%%%%%%%%%%%%%
%%%%%%%%%%%%%%%%%%%%%%%%%%%%

The spectra of quasiparticles consist of 8 branches, 
$\omega=\pm \sqrt{B_1^{\pm}(k)}$ and $\omega=\pm \sqrt{B_2^{\pm}(k)}$. 
They are plotted in Fig.~\ref{fg:disp} for three parameter sets. 
The two flat spectra that appear symmetrically about the abscissa in all cases are 
$\hat\omega=\pm\sqrt{\hat{B}_2^-(\hat{k})}$, which varies only weakly with $\hat{k}$ 
for $\hat{\mu}\gg \hat{V}$.  It is instructive to consider various limits of \eqref{eq:BB}. 
\begin{itemize}
\item When $V\to 0$, 
\ba
	B_1^{\pm}(k) & \to 
		\frac{1}{2}\Big[
		(vk-\mu)^2 +\Delta^2+\mu_\xi^2 \pm 
		\big\{ (vk-\mu)^2+\Delta^2-\mu_\xi^2 \big\}
	\Big]  
	= \begin{cases}(vk-\mu)^2+\Delta^2\,, \\ \mu_\xi^2\,. \end{cases}
	\\
	B_2^{\pm}(k) & \to 
		\frac{1}{2}\Big[
		(vk+\mu)^2 +\Delta^2+\mu_\xi^2 \pm 
		\big\{ (vk+\mu)^2+\Delta^2-\mu_\xi^2 \big\}
	\Big]
	= \begin{cases}(vk+\mu)^2+\Delta^2\,, \\ \mu_\xi^2\,. \end{cases}
\ea
The branches $(vk \pm \mu)^2+\Delta^2$ are just the (squared) 
excitation spectra in superconductors, whereas $\mu_\xi^2$ 
is the (squared) energy level of a free impurity. The gap opening  
$\Delta>0$ corresponds to the level repulsion between 
$\omega=\pm(vk-\mu)$, as can be seen in Fig.~\ref{fg:disp}(a). 
%%%%%%%%%
\item 
When $\Delta\to 0$,
\ba
	B_1^{\pm}(k) & \to \bigg\{ 
		\frac{vk-\mu-\mu_\xi \pm \sqrt{ (vk-\mu+\mu_\xi)^2 + 4V^2 } }{2}
	\bigg\}^2 \,,
	\\
	B_2^{\pm}(k) & \to \bigg\{ 
		\frac{vk+\mu+\mu_\xi \pm \sqrt{ (vk+\mu-\mu_\xi)^2 + 4V^2 } }{2}
	\bigg\}^2 \,. 
\ea
The gap opening $V\ne 0$ stems from the level repulsion between  
$\omega=vk-\mu$ and $\omega=-\mu_\xi$, and between 
$\omega=-vk + \mu$ and $\omega=\mu_\xi$, as seen in 
Fig.~\ref{fg:disp}(b). Physically, it indicates 
the formation of a Kondo singlet, namely the hybridization of conduction bands 
and impurity \cite{Yoshida1966,Read-Newns1983,Coleman1984}. 
The magnetic moment of the impurity is completely screened.  
%%%%%%%%%
\item 
When $\mu_\xi\to 0$, 
\ba
	& B_1^{\pm}(k) \to \bigg\{ 
		\frac{\sqrt{ (vk-\mu)^2 +\Delta^2 + 4V^2 } \pm \sqrt{(vk-\mu)^2 +\Delta^2}}{2}
		\bigg\}^2 \,,
	\label{eq:B001}
	\\
	& B_2^{\pm}(k) \to \bigg\{ 
		\frac{\sqrt{ (vk +\mu)^2 +\Delta^2 + 4V^2 } \pm \sqrt{(vk +\mu)^2 +\Delta^2}}{2}
		\bigg\}^2 \,.
	\label{eq:B002}
\ea
These formulae will be used later.
\end{itemize}

\subsubsection{\label{sc:mf3}Kondo effect at $\Delta=0$ \sc{sc:mf3}}

%%%%%%%%%%%%%%%%%%%%%%%%%%%%
%%%%%%%%%%%%%%%%%%%%%%%%%%%%
\begin{figure*}[bt]
	\centering
%	\raisebox{-2pt}{\includegraphics[width=.37\textwidth]{fg_Vplot}} \qquad \quad 
%	\includegraphics[width=.5\textwidth]{fg_Xi_Deltazero}
%	\put(-483,125){\large $(a)$}
%	\put(-273,125){\large $(b)$}
	\includegraphics[height=.29\textwidth]{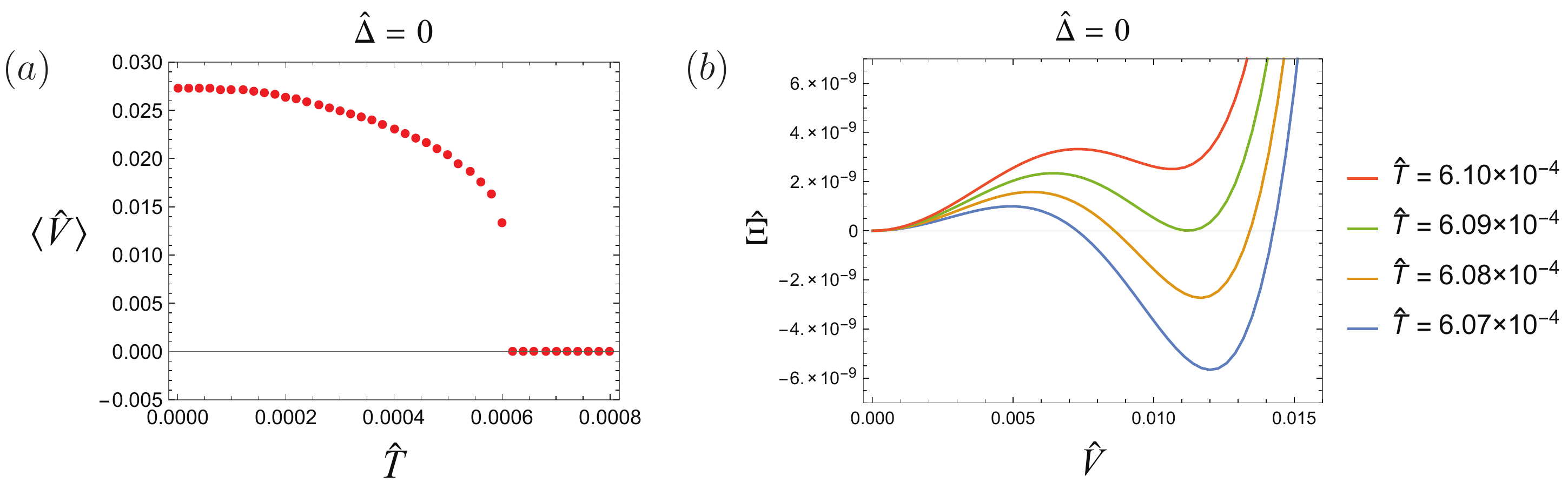}
	\vspace{-6pt}
	\caption{\label{fg:Vplot}(a) Condensate $\langle \hat{V}\rangle$ and (b) 
	thermodynamic potential $\hat\Xi$  
	with $(\hat\mu,~\hat\mu_\xi,~\hat{G})=(0.5,~0,~10)$ 
	at $\hat\Delta = 0$. Here $\hat\Xi$ is normalized to $0$ at the origin.}
\end{figure*}
%%%%%%%%%%%%%%%%%%%%%%%%%%%%
%%%%%%%%%%%%%%%%%%%%%%%%%%%%

Before discussing the effect of the pairing gap on the phase structure, we wish to 
study the Kondo effect at $\Delta=0$ in this model for varying $T$.  In the limit 
$T\to 0$ it is easy to evaluate $\langle V\rangle$ 
in the weak-coupling limit $\hat{G}\ll 1$. This quantity [$V_{T=\Delta=0}(G)$] 
can be determined from the gap equation
\ba
	0 & = \frac{\der \Xi}{\der V^2}\bigg|_{T=\Delta=0} 
	= \frac{1}{G} - \int_0^{\Lambda}\frac{\dd k\,k^2}{2\pi^2} 
	\bigg[
		\frac{1}{\sqrt{ (vk-\mu )^2 + 4V^2 }} + 
		\frac{1}{\sqrt{ (vk+\mu )^2 + 4V^2 }}
	\bigg]\,.
	\label{eq:ewar3}
\ea
Since for small $\hat{G}$ the integral is dominated by 
contributions from the Fermi surface $vk\approx \mu$, $k^2$ in the integrand can be approximated by
$(\mu/v)^2$ and the remaining integral can be done analytically. This leads to%
\footnote{If the number of impurities is finite, there cannot 
be a phase which breaks the $\U(1)$ symmetry of $\xi$ spontaneously. 
In that case the symmetry-breaking condensate $V\ne 0$ observed here 
should be taken as an artifact of the mean-field approximation and $V\to 0$ if we could fully include 
fluctuations around the mean field \cite{Read1985,PhysRevB.35.5072,RevModPhys.59.845}. 
Nevertheless it is worth an emphasis that occurrence of the Kondo effect at $T\sim T_{\rm K}$ 
is well captured by this mean-field method in a qualitatively correct manner.}  
\ba
	V_{T=\Delta=0}(G) \propto \exp \mkakko{ - \frac{\pi^2}{\hat{G}\hat\mu^2}} .
\ea
Thus the spin of the magnetic impurity is screened 
for an \emph{arbitrarily weak} antiferromagnetic interaction $G\to0^+$ at $T=\Delta=0$, 
due to the nonzero Fermi surface \cite{HewsonBook}. 
This property holds regardless of the Dirac nature of the fermions, as long as $\mu>0$. 

By contrast, if the Fermi energy is exactly at the Dirac point ($\mu=0$), the DoS vanishes and the Kondo effect fades away 
unless the interaction is sufficiently strong, i.e., 
$\hat{G}>2\pi^2\simeq 19.74$ in the present model [cf.~Fig.~\ref{fg:ZeroTplots}(b)].%
\footnote{This is conceptually similar to chiral symmetry breaking in the QCD vacuum, where the 
DoS of quarks vanishes but symmetry is broken by strong-coupling effects 
\cite{Nambu:1961tp,*Nambu:1961fr}.} 
While the existence of such a critical coupling at $\mu=0$ 
is consistent with earlier large-$N$ slave-boson analysis of gapless Fermi systems 
\cite{PhysRevLett.64.1835,PhysRevB.46.9274,Borkowski1994,PhysRevB.53.15079,PhysRevB.56.11246,Principi2015}, 
it does not agree with NRG \cite{ChenJayaprakash1995,Ingersent1996,Bulla1997,Ingersent1998} 
according to which the Kondo effect is strictly forbidden for \emph{any} coupling 
in a particle-hole-symmetric Fermi system with the DoS $N(E)\propto|E-E_F|^r$ 
with $r>1/2$. This implies that the Kondo effect in our mean-field analysis at $\mu=0$ 
is an artifact of the approximation used.   
We hasten to add, however, that this criticism only applies to the $\mu=0$ limit and is irrelevant 
for our analysis at $\mu>0$ in the following.

At finite temperature, the IR singularity originating from the sharp Fermi surface is excised and 
the Kondo effect is expected to be suppressed. To see this, we minimize 
the potential \eqref{eq:poten} for each $T$ as a function of $V$. Our 
numerical results are presented in Fig.~\ref{fg:Vplot}, where one observes that 
the Kondo effect is weakened as $T$ increases, and disappears at 
$\hat{T}\simeq 6.09\times 10^{-4}$  
at which the condensate drops to zero through a first-order transition.
By considering that in reality the Kondo effect smoothly emerges as decreasing the temperature,
the first-order transition should be corrected as the crossover.

\subsubsection{\label{sc:mf4}Phase diagram at $\Delta\ne 0$ \sc{sc:mf4}}

%%%%%%%%%%%%%%%%%%%%%%%%%%%%
%%%%%%%%%%%%%%%%%%%%%%%%%%%%
\begin{figure*}[tb]
	\centering
%	\includegraphics[width=.45\textwidth]{fg_Xi_Tempzero} \qquad~
%	\raisebox{-5pt}{\includegraphics[width=.34\textwidth]{fg_phdiag}}
%	\put(-448,123){\large $(a)$}
%	\put(-190,123){\large $(b)$}
	\includegraphics[height=.3\textwidth]{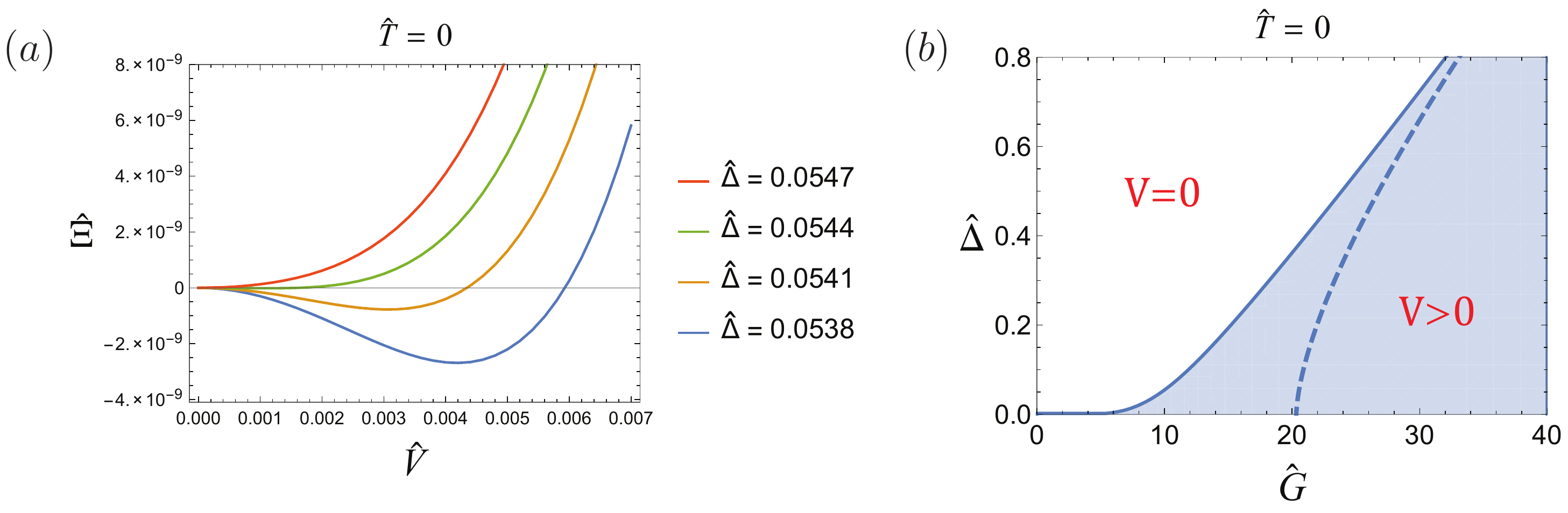}
	\vspace{-9pt}
	\caption{\label{fg:ZeroTplots}
	(a) Thermodynamic potential $\hat\Xi$  
	with $(\hat\mu,~\hat \mu_\xi,~\hat{G})=(0.5,~0,~10)$ at $\hat{T}=0$. 
	The condensate drops to zero for larger $\hat{\Delta}$. 
	(b)~Phase diagram with $(\hat\mu,~\hat\mu_\xi) = (0.5, ~0)$ at $\hat{T}=0$. 
	The Kondo singlet is formed in the shaded region only. 
	The transition across the boundary of the shaded region [$\hat\Delta=2 \hat{V}_{T=\Delta=0}(\hat{G})$] 
	is second order. As soon as $\hat\mu_\xi\ne0$ or $\hat{T}\ne0$, 
	it becomes first order (Fig.~\ref{fg:ZeroTlam}). For comparison, we also 
	show the phase boundary in the limit $\hat\mu\to 0$ (thick dashed line).
	}
\end{figure*}
%%%%%%%%%%%%%%%%%%%%%%%%%%%%
%%%%%%%%%%%%%%%%%%%%%%%%%%%%

\begin{figure}[tb]
	\centering
%	\includegraphics[width=.45\textwidth]{fg_Xi_Tempzero_lambda}
%	\quad 
%	\raisebox{-2pt}{\includegraphics[width=.45\textwidth]{fg_Xi_TempDelta}}
%	\put(-479,121){\large $(a)$}
%	\put(-235,121){\large $(b)$}
	\includegraphics[height=.27\textwidth]{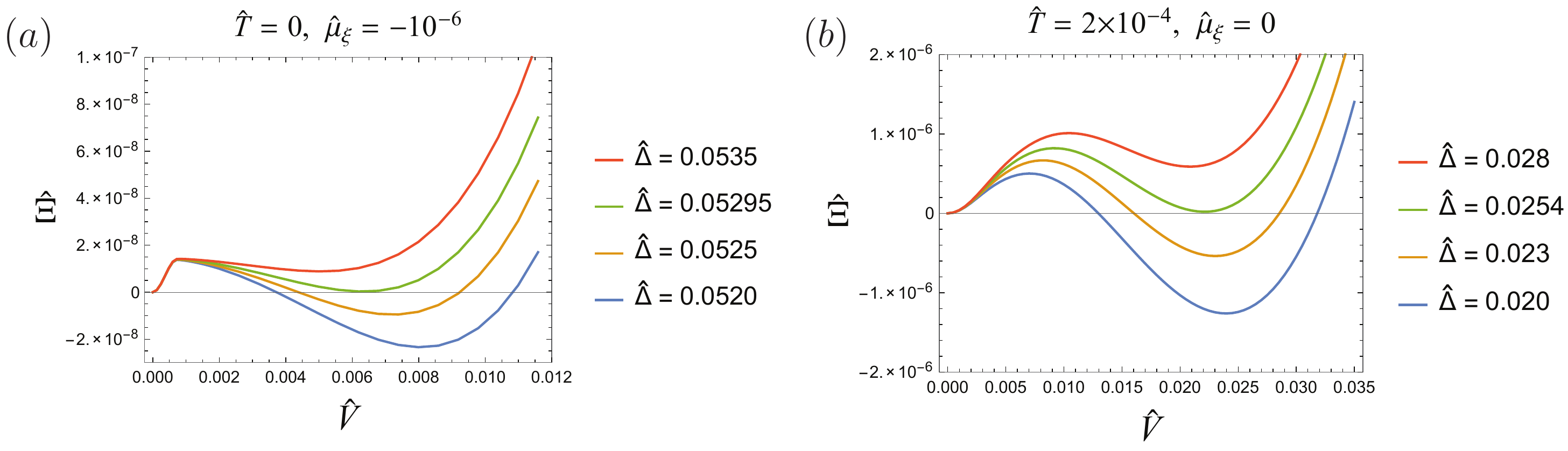}
	\vspace{-3pt}
	\caption{\label{fg:ZeroTlam}
	Thermodynamic potential $\hat\Xi$ with 
	$(\hat\mu,~\hat{G})=(0.5,~10)$. A first-order 
	phase transition occurs as $\hat\Delta$ is increased.  
	}
\end{figure}

\begin{figure*}[tb]
	\centering
	\includegraphics[width=.4\textwidth]{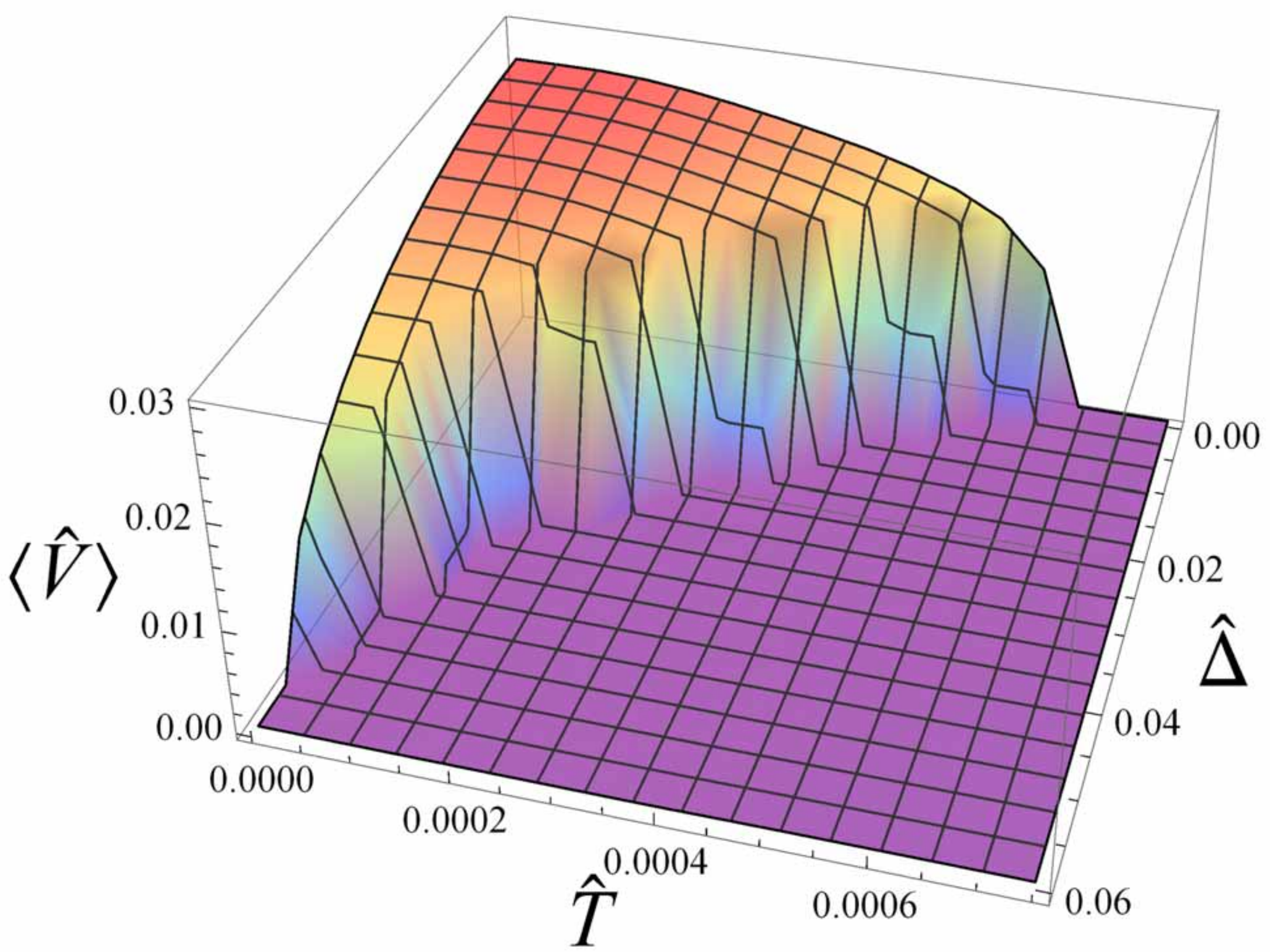}
	\caption{\label{fg:phdiag3d}
	Condensate $\langle \hat{V} \rangle$ with 
	$(\hat\mu,~\hat\mu_\xi,~\hat{G})=(0.5,~0,~10)$. 
	The phase transition is first order except on the $T=0$ axis 
	where it is second order [cf.~Fig.~\ref{fg:ZeroTplots}(a)].
	}
\end{figure*}

We are now prepared to investigate the competition between the Kondo effect 
and superconductivity. We begin with $T=0$, setting $\mu_\xi =0$ 
for simplicity. Plugging \eqref{eq:B001} and \eqref{eq:B002} into \eqref{eq:poten} 
one can easily derive the gap equation  
\ba
	0 & = \frac{1}{G}
	- \int_0^{\Lambda}\frac{\dd k\,k^2}{2\pi^2} 
	\bigg[
		\frac{1}{\sqrt{ (vk-\mu )^2 +\Delta^2 + 4V^2 }} + 
		\frac{1}{\sqrt{ (vk+\mu )^2 +\Delta^2 + 4V^2 }}
	\bigg]\,.
	\label{eq:gap99}
\ea
Comparing \eqref{eq:gap99} with \eqref{eq:ewar3}, one notices an intriguing relation  
\ba
	\langle V \rangle_{T=0} & = \sqrt{ V^2_{T=\Delta=0}(G) - \frac{\Delta^2}{4} }\,,
\ea
which is valid for $0\leq \Delta\leq 2 V_{T=\Delta=0}(G)$. As $\Delta$ increases, 
$\langle V \rangle$ drops continuously and vanishes at $\Delta=2 V_{T=\Delta=0}(G)$.  
For $\Delta>2 V_{T=\Delta=0}(G)$ there is no condensate. 
The second-order nature of this quantum phase transition is clear from 
the evolution of the potential in Fig.~\ref{fg:ZeroTplots}(a). 
In Fig.~\ref{fg:ZeroTplots}(b) we present a phase diagram at $T=0$. It shows that 
for $\Delta>0$, the Kondo screening occurs only when the coupling $G$ is sufficiently strong. 
Thus \emph{the Kondo effect is suppressed by a pairing gap}, confirming our statements 
in Sec.~\ref{sc:os}. We remark that the same picture has been obtained 
long time ago concerning gapped Fermi systems 
\cite{Saso1992,Takegahara1992,Ogura1993,ChenJayaprakash1998} 
and non-Dirac fully gapped superconductors 
\cite{Soda01091967,Satori1992,Sakai1993}.     
In \cite{Soda01091967} it was shown with a diagrammatic method, that  
unlike in normal metals where the antiferromagnetic coupling grows logarithmically 
$\sim \log T$ as $T\to 0$, it saturates at $\sim \log \Delta$ in superconductors, reflecting that the 
Fermi-surface effect is cut off by $\Delta$. We expect that an analogous perturbative calculation 
for a Dirac superconductor would be able to confirm our conclusion obtained in the 
mean-field approach.   
Summarizing, the ground state of a magnetic impurity at zero temperature is expected to be 
\ba
	\label{eq:doubletsinglet}
	\begin{array}{rll}
		\star & \text{a singlet} ~~& \text{for}~~ 
		0\leq \Delta < 2 V_{T=\Delta=0}(G)\,, 
		\vspace{5pt}
		\\
		\star & \text{a doublet} ~~&
		\text{for}~~ \Delta > 2 V_{T=\Delta=0}(G)\,. 
	\end{array}
\ea
The second-order phase transition found above immediately turns into 
\emph{first order} as soon as we switch on small $\mu_\xi\ne0$ or 
$T\ne 0$, as illustrated in Fig.~\ref{fg:ZeroTlam}. Therefore the transition between 
the two states \eqref{eq:doubletsinglet} is generically first order, as was 
already emphasized in Fig.~\ref{fg:OSp}. 
%%%%%%%%%%%%%%%%%%%%%%%%%%%%
%%%%%%%%%%%%%%%%%%%%%%%%%%%%
%%%%%%%%%%%%%%%%%%%%%%%%%%%%
%%%%%%%%%%%%%%%%%%%%%%%%%%%%
Based on investigations so far, one can map out the full phase diagram with both $\Delta\ne 0$ and $T\ne 0$. 
The numerically obtained condensate $\langle \hat{V} \rangle$ is plotted in Fig.~\ref{fg:phdiag3d}.  

%%%%%%%%%%%%%%%%%%%%%%%%%%%%
%%%%%%%%%%%%%%%%%%%%%%%%%%%%
%%%%%%%%%%%%%%%%%%%%%%%%%%%%
%%%%%%%%%%%%%%%%%%%%%%%%%%%%

\begin{figure*}[tb]
	\centering
	\includegraphics[height=.18\textwidth]{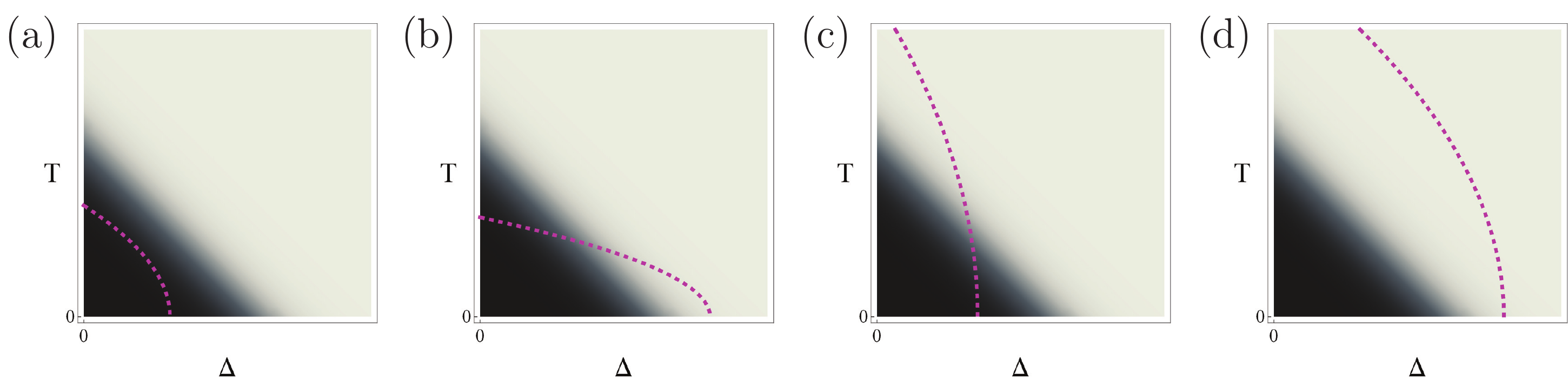}
	\caption{\label{fg:Deltacontour_1_2_3_4}
	Schematic plots for the temperature dependence of $\Delta$ in real materials. 
	The dashed line denotes the curve for $\Delta(T)$ while the dark area indicates 
	the domain with a strong Kondo effect. Which of these possibilities is realized depends 
	on microscopic details of the material and the impurity. The back reaction of impurities 
	on the superconducting gap is neglected, assuming a sufficiently low density of impurities. 
	}
\end{figure*}

Although we have treated the Majorana mass $\Delta$ as a parameter in the above discussions, 
the gap in actual superconductors is dynamically determined 
and hence depends on temperature. The gap function $\Delta=\Delta(T)$ determines 
a curve on the $(\Delta,T)$-plane. In Fig.~\ref{fg:Deltacontour_1_2_3_4} we overlaid 
several possible behaviors of the gap function on the phase diagram of the Kondo impurity.  
When $T_{\rm K}$ exceeds $\Delta(0)$, the Kondo effect would persist throughout the 
superconducting phase [Case (a)]. Conversely, when $\Delta(0)$ exceeds $T_{\rm K}$, there would be no 
temperature region in which the Kondo effect survives [Case (d)].  Intermediate cases (b) and (c) are also 
possible when $T_{\rm K}$ and $\Delta(0)$ are numerically close.  

%%%%%%%%%%%%%%%%%%%%%%%%%%%%
%%%%%%%%%%%%%%%%%%%%%%%%%%%%
%%%%%%%%%%%%%%%%%%%%%%%%%%%%
%%%%%%%%%%%%%%%%%%%%%%%%%%%%

\subsection{\label{sc:sbs1}Shiba states \sc{sc:sbs1}}

The previous sections dealt with ground state properties. In this 
section, we turn to the analysis of excited states. 
Magnetic impurities in superconductors are known to 
induce localized excited states in the gap \cite{RevModPhys.78.373}. 
In the case of a conventional $s$-wave superconductor, 
Shiba \cite{Shiba01091968} demonstrated this by solving a model of a 
classical spin immersed in a superconductor to all orders in the interaction, where 
``classical'' means that the impurity was treated as a spin-dependent external potential. 
The picture that emerged from \cite{Shiba01091968} was qualitatively 
consistent with elaborate NRG studies \cite{Satori1992,Sakai1993} 
across the whole range of $T_{\rm K}/T_{c}$ ($T_c$: the superconducting transition 
temperature), including the presence of a quantum phase transition at an intermediate coupling. 
Similar analysis was performed also for anisotropic superconductors \cite{Borkowski1994,PhysRevB.55.12648}.  
By contrast, the slave-boson mean-field theory yields rather inaccurate predictions for the Shiba state 
especially in the regime $T_{\rm K}\lesssim T_c$ \cite{Borkowski1994}, 
because the slave-boson theory is not capable of 
describing correlations at temperatures $\gtrsim T_{\rm K}$ where the slave-boson 
expectation value vanishes. This is particularly serious when studying dense quark 
matter where $T_{\rm K}/T_c \sim T_{\rm K}/\Delta \ll 1$ 
(cf.~Sec.~\ref{sc:os}).  For this reason, we will in the following 
discuss properties of the Shiba state associated with a \emph{classical} 
impurity in a Dirac superconductor employing the $T$-matrix method of \cite{Shiba01091968}.  

We consider a Hamiltonian of a superconducting $N_f=1$ 
Weyl fermion in the presence of a localized classical impurity, 
\ba
	\hat{H} & \equiv \int \dd^3x \kkakko{
		\psi^\dagger (
			iv \bm{\sigma\cdot} \nabla - \mu 
		) \psi  
		- \Delta \big(
		\psi_{\up} \psi_{\down} + 
		\psi_{\down}^{\dagger}\psi_{\up}^{\dagger}
		\big) 
		+ \psi^\dagger U(\mathbf{x}) \psi 
	}. 
	\label{eq:Hclassical}
\ea
The last term represents a potential due to the magnetic impurity. If $U(\mathbf{x})$ is 
independent of the spin, it is just a potential scattering and will not be 
of interest to us. Let us take $U(\mathbf{x})=u\,\delta(\mathbf{x})\sigma^3$ 
where $u$ is a parameter characterizing the strength of the potential; we 
assume $u>0$ without loss of generality. (The sign of $u$ can be flipped by 
a spin rotation.) In \eqref{eq:Hclassical} the gap $\Delta$ is assumed to be 
spatially uniform, though $\Delta$ may slightly vary near the impurity site 
\cite{Sakai1993}.

\begin{figure}[bt]
	\centering
	\includegraphics[width=10cm]{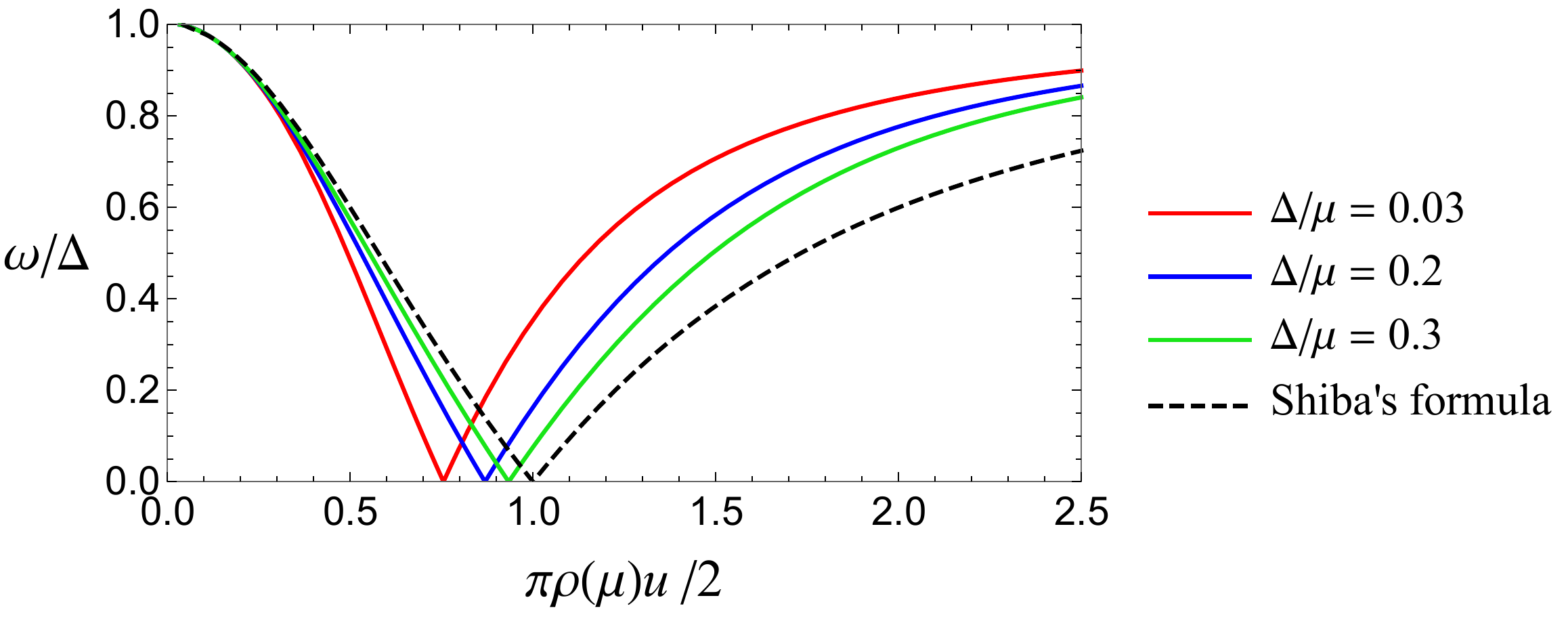}
	\vspace{-5pt}
	\caption{\label{fg:levelcrossing}Spectrum of mid-gap excited states 
	numerically obtained from \eqref{eq:4646} with $v\Lambda/\mu=2$, 
	in comparison with Shiba's formula \eqref{eq:omful}. 
	Negative energy levels (not shown) appear symmetrically about the horizontal axis.
	}
\end{figure}

The task is to obtain the mid-gap excitation level 
with the $T$-matrix method. In the Nambu basis 
$\npsi_{\bk}\equiv\big( \psi_{\bk\up} \quad   
\psi_{\bk\down} \quad  \psi^\dagger_{-\bk\up} \quad   
\psi^\dagger_{-\bk\down} \big)^{\rm T}$, 
the Hamiltonian without an impurity reads  
%\ba
%	& \frac{1}{2}\int \frac{\dd^3\bk}{(2\pi)^3}
%	\npsi^\dagger_\bk
%	\bep
%		-vk_3-\mu&-v(k_1-ik_2)&0&\Delta
%		\\
%		-v(k_1+ik_2)&vk_3-\mu&-\Delta&0
%		\\ 
%		0&-\Delta&-vk_3+\mu&-v(k_1+ik_2)
%		\\
%		\Delta&0&-v(k_1-ik_2)&vk_3+\mu
%	\eep
%	\npsi_\bk 
%	\equiv 
%	\frac{1}{2}\int \frac{\dd^3\bk}{(2\pi)^3}
%	\npsi^\dagger_\bk
%	\hat{\cal H}_\bk 
%	\npsi_\bk \,,
%\ea
\ba
	& \frac{1}{2}\int \frac{\dd^3 k}{(2\pi)^3}
	\npsi^\dagger_\bk
	\bep
		-v \bk\bm{\cdot\sigma} - \mu & i\sigma^2 \Delta 
		\\
		-i\sigma^2\Delta & -v \bk\bm{\cdot\sigma}^{\rm T}+\mu 
	\eep
	\npsi_\bk 
	\equiv 
	\frac{1}{2}\int \frac{\dd^3 k}{(2\pi)^3}
	\npsi^\dagger_\bk
	\hat{\cal H}_\bk 
	\npsi_\bk \,,
\ea
where we have omitted terms that arise from the anticommutator of 
$\psi_{-\bk}$ and $\psi^\dagger_{-\bk}$. 
The eigenvalues of $\hat{\cal H}_\bk$ are given by 
$\pm\sqrt{(vk-\mu)^2+\Delta^2}$ and 
$\pm \sqrt{(vk+\mu)^2+\Delta^2}$ with $k\equiv |\bk|$. 
The Green's function of a clean system is defined as
\ba
	\hat{G}_0(\bk,\omega) = \frac{1}{\omega - \hat{\cal H}_\bk}\,.
\ea
Since $\hat{\cal H}_\bk$ has a spectral gap at $[-\Delta,\Delta]$, 
$\hat{G}_0(\bk,\omega)$ has no pole at $|\omega|< \Delta$. 
The impurity part reads
\ba
	\int\dd^3x~\psi^\dagger U(\mathbf{x}) \psi 
%	& = \int\dd^3x~\psi^\dagger 
%	u \, \delta(\mathbf{x})\sigma^3 \psi
%	\\
%	& = u \int \frac{\dd^3\bk}{(2\pi)^3}
%	\int \frac{\dd^3\bk'}{(2\pi)^3}
%	\psi^\dagger_{\bk'} \sigma^3 \psi_\bk
%	\\
	& = \frac{1}{2}\int \frac{\dd^3 k}{(2\pi)^3}
	\int \frac{\dd^3 k'}{(2\pi)^3}
	\npsi^\dagger_{\bk'}
	\hat{U} \npsi_\bk  
	\quad\qquad \text{with}~~~
	\hat{U} \equiv \bep u\sigma^3 & 0\\0&-u\sigma^3 \eep .
\ea
Averaging over the impurity distribution (which restores 
translational invariance), one obtains the full Green's function 
in terms of the $T$-matrix $\hat{T}(\omega)$ as 
\cite{Mahan2000,Altland:2006si} 
\ba
	\hat{G}(\bk,\omega) & =  
	\hat{G}_0(\bk,\omega) 
	+ n_i \hat{G}_0(\bk,\omega)\hat{T}(\omega)
	\hat{G}_0(\bk,\omega) 
	+ \calO(n_i^2)\,,
\ea
where $n_i$ is the impurity density and 
\ba
	\hat{T}^{-1}(\omega) & = \hat{U}^{-1} 
	- \int \frac{\dd^3 k}{(2\pi)^3} \hat{G}_0(\bk,\omega) \,.
	\label{eq:ttt}
\ea
The impurity-induced mid-gap state 
manifests itself as a pole of $\hat{T}(\omega)$ 
at $|\omega|<\Delta$. Thus the equation to be solved is
\ba
	\det\big[ \hat{T}^{-1}(\omega) \big] 
	= \det\kkakko{ 
		\hat{U}^{-1} 
		- \int \frac{\dd^3 k}{(2\pi)^3} \hat{G}_0(\bk,\omega)
	} 
	\overset{!}{=} 0\,.
\ea
The integral is evaluated as
\ba
	\scalebox{0.87}{$\displaystyle 
		\int \frac{\dd^3 k}{(2\pi)^3} 
		\hat{G}_0(\bk,\omega) 
		= \int \frac{\dd^3 k}{(2\pi)^3} 
		\frac{1}{\Y^2-4\mu^2v^2k^2}
		\bep
			(\omega-\mu)\Y - 2\mu v^2k^2 &0&0& \Delta \Y
			\\
			0& (\omega-\mu)\Y - 2\mu v^2k^2 & - \Delta \Y &0
			\\
			0& - \Delta \Y & (\omega+\mu) \Y + 2\mu v^2k^2  &0
			\\
			\Delta \Y &0&0& (\omega+\mu) \Y + 2\mu v^2k^2 
		\eep
	$}, 
	\label{eq:Gmatr}
\ea
where $\Y \equiv \omega^2-\mu^2-\Delta^2-v^2k^2$. 
The $4\times 4$ matrix above is a direct sum of two $2\times 2$ blocks. For simplicity, 
we will hereafter focus on the central $2\times 2$ block of \eqref{eq:Gmatr}. Then
\ba
	\scalebox{0.9}{$\displaystyle 
		\hat{T}^{-1}(\omega) = -u^{-1}\1_2 - 
		\int^{\Lambda}\!\!\! \frac{\dd^3 k}{(2\pi)^3} 
		\frac{1}{
			[(vk-\mu)^2+\Delta^2-\omega^2]
			[(vk+\mu)^2+\Delta^2-\omega^2]
		}
		\bep
			(\omega-\mu)\Y - 2\mu v^2k^2 & - \Delta \Y 
			\\
			- \Delta \Y & (\omega+\mu)\Y + 2\mu v^2k^2
		\eep .
	$}
	\label{eq:4646}
\ea
The divergent momentum integral was regularized by a cutoff $\Lambda$. 
If a weak superconductor $\Delta\ll \mu$ is considered, 
$\Y \approx -\mu^2-v^2k^2$. If, furthermore, 
we restrict the domain of integration to a thin shell around 
the Fermi momentum as in the conventional BCS theory, then 
$\Y \approx -2\mu^2$ and 
\ba
	\hat{T}^{-1}(\omega) 
	& \approx -u^{-1}\1_2 - \!\!
	\int\limits_{|vk-\mu|<\delta}\!\!\! \frac{\dd^3 k}{(2\pi)^3} 
	\frac{1}{
		[(vk-\mu)^2+\Delta^2-\omega^2]
		\cdot 4\mu^2
	}
	\bep
		-2 \mu^2 \omega & 2\mu^2\Delta  
		\\
		2\mu^2\Delta & - 2\mu^2 \omega 
	\eep 
	\\
	& = -u^{-1}\1_2 - \frac{1}{2}\frac{1}{(2\pi)^3}\frac{4\pi\mu^2}{v^3}
	\int\limits_{\mu-\delta}^{\mu+\delta} \dd E\ 
	\frac{1}{
		(E-\mu)^2+\Delta^2-\omega^2
	}
	\bep
		- \omega & \Delta  
		\\
		\Delta & - \omega 
	\eep 
	\qquad [E\equiv vk]
	\\
	& \approx -u^{-1}\1_2 - \frac{\pi \rho(\mu)}{2}
	\frac{1}{\sqrt{\Delta^2-\omega^2}}
	\bep
		- \omega & \Delta  
		\\
		\Delta & - \omega 
	\eep\,,
	\label{eq:Tishiba}
\ea
where in the last step the domain of integration over $E$ was extended to $\pm\infty$ to allow for exact 
integration. We have introduced the DoS at the Fermi energy in a normal phase, 
$\rho(\mu) \equiv \mu^2/(2\pi^2v^3)$. Now the equation $\det[\hat{T}^{-1}(\omega)]=0$ can be solved 
analytically and yields the energy of the Shiba bound state, which 
resides in the gap $[-\Delta,\Delta]$ for an \emph{arbitrarily weak} interaction: 
\ba
	\frac{\omega}{\Delta} = 
	\frac{1-[\pi \rho(\mu)u/2]^2}{1+[\pi \rho(\mu)u/2]^2}\,.
	\label{eq:midg}
\ea
This expression exactly agrees with Shiba's formula \cite{Shiba01091968} 
derived for fermions with a quadratic dispersion. Notably, the RHS of 
\eqref{eq:midg} does not depend on $\Delta$. No such mid-gap pole 
is found in the case of a spin-independent potential 
[$U({\bf x})\propto \1_2]$. Equation \eqref{eq:midg} (plotted in 
Fig.~\ref{fg:levelcrossing} as ``Shiba's formula'') 
shows that the excitation energy becomes negative 
when the interaction exceeds 
$\pi \rho(\mu)|u|/2=1$, indicative of a level crossing between the screened 
and unscreened states. 
This is a manifestation of the first-order phase transition observed in Sec.~\ref{sc:mf}. 
(Making $u$ larger corresponds to making $\Delta$ smaller at fixed coupling in the setup of Sec.~\ref{sc:mf}.) 
The bound state energy lies close to the gap edges 
both for strong $[\rho(\mu)|u|\gg 1]$ and weak $[\rho(\mu)|u|\ll 1]$ couplings. 
We remark that the same result \eqref{eq:midg} but with an overall \emph{minus sign} 
follows had we considered the other $2\times 2$ sub-block of \eqref{eq:Gmatr}. 
Thus there are two mid-gap states altogether whose energies are%
\footnote{The Shiba states in the present classical treatment do not distinguish 
$u>0$ from $u<0$. Once quantum effects of the impurity spin are taken into account, 
the difference between a ferromagnetic/antiferromagnetic 
coupling becomes qualitatively important \cite{RevModPhys.78.373}.} 
\ba
	\frac{\omega}{\Delta} = \pm 
	\frac{1-[\pi \rho(\mu)u/2]^2}{1+[\pi \rho(\mu)u/2]^2}\,. 
	\label{eq:omful}
\ea

%%%%%%%%%%%%%%%%%%%%%%%%%%%%%%%%%%%%%%
%%%%%%%%%%%%%%%%%%%%%%%%%%%%%%%%%%%%%%

To estimate the energy of the Shiba states without resorting to the thin-shell approximation for the momentum 
integral, we have numerically evaluated \eqref{eq:4646} directly and compared 
the resulting energy with Shiba's formula, as shown in Fig.~\ref{fg:levelcrossing}. 
(Only $\omega/\Delta\geq 0$ is displayed.) The global trend is the same for all values of 
$\Delta/\mu$, but the locus of the transition differs slightly. 
All the curves coincide for $\pi\rho(\mu)|u|/2\lesssim 0.3$, 
indicating that the asymptotic form $\omega/\Delta \simeq \pm(1-2[\pi \rho(\mu)u/2]^2)$ 
at weak coupling is universal and independent of the UV regularization and $\Delta/\mu$.

\section{\label{sc:cicsq}Impurity in color superconductors \sc{sc:cicsq}}

\subsection{\label{sc:sqk}Mean-field theory \sc{sc:sqk}}
\subsubsection{\label{sc:sqk1}Model setup \sc{sc:sqk1}}

In this section we wish to gain insights into the competition between 
the Kondo effect and color superconductivity in the 2SC phase, which is 
one of the candidate phases of QCD that may be realized in the interior of compact stars 
\cite{Alford:1997zt,Rapp:1997zu}.  
If the gapless quarks of third color that are neutral under the 
unbroken $\SU(2)$ gauge group decouple, then one may adopt a toy model 
of two-color two-flavor Dirac fermions with a Majorana mass $\Delta$ as a crude but 
reasonable approximation to the 2SC phase. For technical simplicity we limit 
ourselves to the sector of a single chirality of quarks, which leads to the model 
\ba
	S & = \int \dd\tau\,\dd^3x \bigg[
		\sum_{f,a} 
		\psi_{fa}^\dagger (\der_{\tau} - \mu + i v \bm{\sigma\cdot}\nabla) \psi^{}_{fa} 
		+ 
		\frac{\Delta}{2} (\psi^\text{T}\sigma^2 \tau^2 t^2 \psi + 
		\psi^{\dagger}\sigma^2 \tau^2 t^2 \psi^*)  
		+ \sum_{a,s} \xi_{as}^\dagger (\der_\tau - \mu_\xi) \xi_{as}  
	\notag
	\\
	& \qquad \qquad \qquad 
		+ \frac{G}{4} \sum_{f} \sum_{\mu,A=0}^{3} 
		(\psi_f^\dagger \sigma^\mu t^A \psi_f)(\xi^\dagger \sigma^\mu t^A \xi) 
	\bigg]. 
	\label{eq:lagw}
\ea
Here $s=1,2$, $f=1,2$, and $a=1,2$ are spins, flavors and colors, respectively. 
$\sigma^{1,2,3}$, $\tau^{1,2,3}$ and $t^{1,2,3}$ are Pauli matrices 
in the spin, flavor, and color space, respectively, and $\sigma^0=t^0=\1_2$. 
An important distinction from the model \eqref{eq:mdl} considered in Sec.~\ref{sc:os} 
is that the impurity in \eqref{eq:lagw} has spin, in addition to color. 
Our model is similar to the two-color two-flavor Nambu--Jona-Lasinio (NJL) model 
\cite{Ratti:2004ra}.% 
\footnote{The dynamics of two-color QCD sensitively depends on the number of 
flavors. For an even number of flavors the phase structure at finite density is 
relatively well understood \cite{Kogut:2000ek,Kogut:2000ek,Splittorff:2000mm,
Sun:2007fc,Kanazawa:2009ks}. The case of odd flavors remains less explored.} 
A Kondo-like model with an interaction analogous to \eqref{eq:lagw} was studied 
in \cite{Pang1994,Kuramoto1998}.  The gap $\Delta$ 
breaks the $\U(1)$ symmetry of $\psi$ while preserving the $\SU(2)$ color and $\SU(2)$ 
flavor symmetry.  These two groups are interchangeable, for 
the color $\SU(2)$ is realized as a \emph{global} symmetry in this model. 

The interaction in \eqref{eq:lagw} with $G>0$ models the color-current interaction 
mediated by gluons in QCD. It comprises four pieces,
\ba
	\sum_{\mu,A=0}^{3} 
	(\psi_f^\dagger \sigma^\mu t^A \psi_f^{})(\xi^\dagger \sigma^\mu t^A \xi) 
	& = 
	(\psi_f^\dagger\psi_f^{})(\xi^\dagger\xi) + 
	\sum_{i=1}^{3} 
	(\psi_f^\dagger \sigma^i \psi_f^{})(\xi^\dagger \sigma^i \xi)  
	+ \sum_{j=1}^{3} (\psi_f^\dagger t^j \psi_f^{})(\xi^\dagger t^j \xi)  
	+ \sum_{i,j=1}^{3} 
	(\psi_f^\dagger \sigma^i t^j \psi_f^{})(\xi^\dagger \sigma^i t^j \xi)\,.
	\label{eq:intint}
\ea
It is a sum of potential scattering, spin-exchange interaction,%
\footnote{Strictly speaking, heavy quarks in quark matter do not experience 
spin-dependent interactions 
\cite{Isgur:1989vq,*Isgur:1989ed,Neubert:1993mb,Manohar2000Book}.} 
color-exchange interaction, and spin-color exchange interaction. 
In principle the coupling for each term may be chosen independently, though 
they mix under renormalization \cite{Kuramoto1998}. The model \eqref{eq:lagw} 
makes all couplings equal to make the mean-field analysis easier.  
The exact screening of the impurity amounts to the screening of both color and spin. 
This is achieved by a single flavor of $\psi$. As there are two flavors of $\psi$ 
in the model \eqref{eq:lagw} the impurity will be subject to 
the \emph{overscreened} Kondo effect at $\Delta=0$.

Using $\sum_{\mu=0}^{3}(\sigma^\mu)_{ij}(\sigma^\mu)_{kl}=
\sum_{A=0}^{3}(t^A)_{ij}(t^A)_{kl}=2\delta_{il}\delta_{jk}$ 
for the interaction part of \eqref{eq:lagw}, one obtains 
\ba
	S & =\int \dd^4x \bigg[
		\psi_{fa}^\dagger (\der_\tau - \mu+i v \bm{\sigma\cdot}\nabla) \psi_{fa} 
		+ \frac{\Delta}{2} (\psi^\text{T}\sigma^2 \tau^2 t^2 \psi + 
		\psi^{\dagger}\sigma^2 \tau^2 t^2 \psi^* ) 
		+ \xi^\dagger_{as} (\der_\tau - \mu_\xi) \xi_{as}
		- G (\psi_{fas}^\dagger \xi_{as}) (\xi_{a's'}^\dagger \psi^{}_{fa's'})    
	\bigg] 
\ea
where repeated indices are summed. 
Next, we perform a Hubbard-Stratonovich transformation with a color-singlet auxiliary field $V_f$,
\ba
	- G (\psi_f^\dagger \xi) (\xi^\dagger \psi_f) 
	&~ \Rightarrow~ \frac{|V_f|^2}{G} + V_f^* \psi^\dagger_f \xi + V_f \xi^\dagger \psi_f
\ea
and apply a mean-field approximation, $V_1=\text{const.}$ and $V_2=\text{const.}$ 
Without loss of generality, one can rotate the mean field in the flavor space 
so that only the first flavor mixes with the impurity: 
\begin{gather}
	\label{eq:mixing2sc}
	\mathfrak{g}\bep V_1 \\ V_2 \eep = \bep \sqrt{|V_1|^2+|V_2|^2} \\ 0 \eep 
	\equiv \bep V \\ 0 \eep 
	\qquad \text{with}\quad 
	\mathfrak{g} \equiv \frac{1}{\sqrt{|V_1|^2+|V_2|^2}}
	\bep V_1^* & V_2^* \\ - V_2 & V_1 \eep \in \SU(2)\,.
\end{gather}
This pattern of hybridization is indicative of the ASC phase (Sec.~\ref{sc:os}, 
Table~\ref{tb:phases}), i.e., the ground state at $V\ne 0$ is a flavor doublet with no 
color and spin. We will come back to this point later.
The values of $V$ and $\mu_\xi$ should be determined by solving \eqref{eq:Xiextrema}. 

\begin{figure*}[bt]
	\centering
%	\includegraphics[width=.24\textwidth]{fg_disp_2sc_A} \qquad \quad
%	\includegraphics[width=.24\textwidth]{fg_disp_2sc_B} \qquad \quad
%	\includegraphics[width=.24\textwidth]{fg_disp_2sc_C} 
%	\put(-443,115){\large $(a)$}
%	\put(-288,115){\large $(b)$}
%	\put(-137,115){\large $(c)$}
	\includegraphics[height=.26\textwidth]{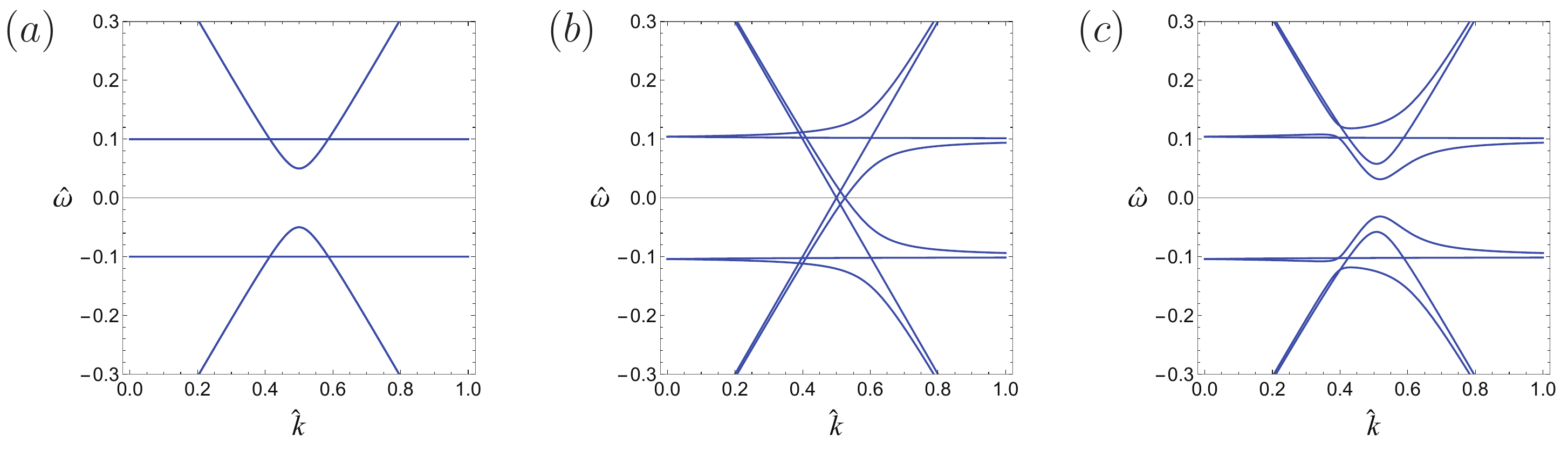} 
	\vspace{-.5\baselineskip}
	\caption{\label{fg:disp2sc}Dispersion relations of quasiparticles 
	$\{\hat\omega = \pm \hat E_{\ell}(\hat k),~\ell=1,\dots,6 \}$  
	for $\hat\mu=0.5$ and $\hat\mu_\xi=-0.1$.  
	(a):~$(\hat\Delta,\hat{V})=(0.05, 0)$, (b):~$(\hat\Delta,\hat{V})=(0, 0.05)$, 
	(c):~$(\hat\Delta,\hat{V})=(0.05, 0.05)$. }
\end{figure*}

In the Nambu basis, the action reads%
\footnote{A caution concerning \eqref{eq:NGbase}: 
$\1_4=\1_2^{\rm color}\otimes\1_2^{\rm flavor}$ for $\psi$ 
and $\1_4=\1_2^{\rm color}\otimes\1_2^{\rm spin}$ for $\xi$.}  
\ba
	\scalebox{0.9}{$\displaystyle 
	S = \int \dd^4 x \bigg[ \frac{1}{2}
	\bep \psi^\text{T} & \psi^\dagger & \xi^\text{T} & \xi^\dagger \eep
	\bep
		\Delta \sigma^2 \tau^2 t^2  & 
		(\der_\tau+\mu+iv \bm{\sigma}^{\rm T}\!\bm{\cdot}\nabla)\otimes \1_4  & 0 & - \hat{V}(4)^{\rm T}
		\\
		(\der_\tau-\mu+iv \bm{\sigma\cdot}\nabla)\otimes \1_4 & \Delta \sigma^2 \tau^2 t^2 & \hat{V}(4)^{\rm T} & 0
		\\
		0 &-\hat{V}(4) &0& (\der_\tau + \mu_\xi) \1_4
		\\
		\hat{V}(4) & 0 & (\der_\tau - \mu_\xi) \1_4 &0
	\eep
	\bep \psi \\ \psi^* \\ \xi \\ \xi^* \eep
	+ \frac{V^2}{G} \bigg]\,,
	$}
	\label{eq:NGbase}
\ea
where $\displaystyle \hat{V}(n) \equiv \big( V\1_n \quad 0 \1_n \big)$. 
(Recall that $\psi$ has a flavor index but $\xi$ does not.) 
The thermodynamic potential per unit volume is obtained as 
\ba
	\scalebox{0.95}{$\displaystyle 
	\Xi = \frac{V^2}{G} - \frac{1}{2} T \sum_{n\in\ZZ} \int\frac{\dd^3k}{(2\pi)^3}
	\log \det
	\bep
		\Delta \sigma^2 \tau^2 t^2  & 
		(i\omega_n+\mu - v \bm{\sigma}^{\rm T}\!\bm{\cdot k})\otimes \1_4  & 
		0 & - \hat{V}(4)^{\rm T}
		\\
		(i\omega_n-\mu - v \bm{\sigma\cdot k})\otimes \1_4 & 
		\Delta \sigma^2 \tau^2 t^2 & \hat{V}(4)^{\rm T} & 0
		\\
		0 &-\hat{V}(4) &0& (i\omega_n + \mu_\xi) \1_4
		\\
		\hat{V}(4) & 0 & (i\omega_n - \mu_\xi) \1_4 &0
	\eep .
	$} \!\!
\ea
Let us denote this $24\times24$ matrix by $\bigstar$. The structure 
of $\bigstar$ can be simplified by noticing that $t^2$ can be diagonalized as 
$\tilde{\mathfrak{g}}^\dagger t^2 \tilde{\mathfrak{g}}
=t^3$ with $\tilde{\mathfrak{g}}\equiv (\1_2+it^1)/\sqrt{2}$. Then 
$\bigstar$ becomes diagonal in the color space and a quick inspection shows 
\ba
	\det \bigstar  &  = (\det \star) \times (\det \star \big|_{\Delta\to -\Delta})
	= (\det \star)^2  \,,
\ea
where 
\ba
	\star \equiv \bep
		\Delta \sigma^2 \tau^2  & 
		(i\omega_n+\mu - v \bm{\sigma}^{\rm T}\!\bm{\cdot k})\otimes \1_2  & 
		0 & - \hat{V}(2)^{\rm T}
		\\
		(i\omega_n-\mu - v \bm{\sigma\cdot k})\otimes \1_2 & 
		\Delta \sigma^2 \tau^2  & \hat{V}(2)^{\rm T} & 0
		\\
		0 &-\hat{V}(2) &0& (i\omega_n + \mu_\xi) \1_2
		\\
		\hat{V}(2) & 0 & (i\omega_n - \mu_\xi) \1_2 &0
	\eep .
\ea
Its determinant can be factored as 
$\displaystyle (\det\star)=\prod_{\ell=1}^{6}[\omega_n^2+E_{\ell}(k)^2]$ with 
$E_\ell(k)\geq 0$. The explicit formulas for $E_\ell(k)$ are quite lengthy and we refrain from reproducing them here.%
\footnote{We used \emph{Mathematica 10.4} \cite{Mathematica104} to solve the equations for $E_\ell$.} 
In Fig.~\ref{fg:disp2sc} we display the quasiparticle dispersion relations. (The notation follows \eqref{eq:hatnotation}.) 
The obtained spectrum is essentially a superposition of Fig.~\ref{fg:disp} on top of a free dispersion 
with a gap $\Delta$. The latter represents the second flavor that does not hybridize with the impurity 
[cf.~\eqref{eq:mixing2sc}]. 

%%%%%%%%%%%%%%%%%%%%%%%%%%%%
%%%%%%%%%%%%%%%%%%%%%%%%%%%%
%%%%%%%%%%%%%%%%%%%%%%%%%%%%
%%%%%%%%%%%%%%%%%%%%%%%%%%%%

Proceeding as before, we find the thermodynamic potential  
\ba
	\Xi & = \frac{V^2}{G} - T \sum_{n\in\ZZ} 
	\sum_{\ell=1}^{6}\int\frac{\dd^3k}{(2\pi)^3}
	\log[\omega_n^2+E^2_\ell(k)]
	\\
	& = \frac{V^2}{G} - T \sum_{\ell=1}^{6}
	\int_0^{\Lambda}\frac{\dd k\,k^2}{\pi^2}
	\log \cosh \frac{E_\ell(k)}{2T}\,.
\ea

\subsubsection{\label{sc:sqk2}Numerical results \sc{sc:sqk2}}

\begin{figure}[tb]
	\centering
%	\raisebox{4pt}{\includegraphics[width=.45\textwidth]{fg_Xi_Delta_2sc}} 
%	\qquad
%	\includegraphics[width=.4\textwidth]{fg_phdiag3D_V_2sc} 
%	\put(-460,130){\large $(a)$}
%	\put(-207,130){\large $(b)$}
	\includegraphics[height=.3\textwidth]{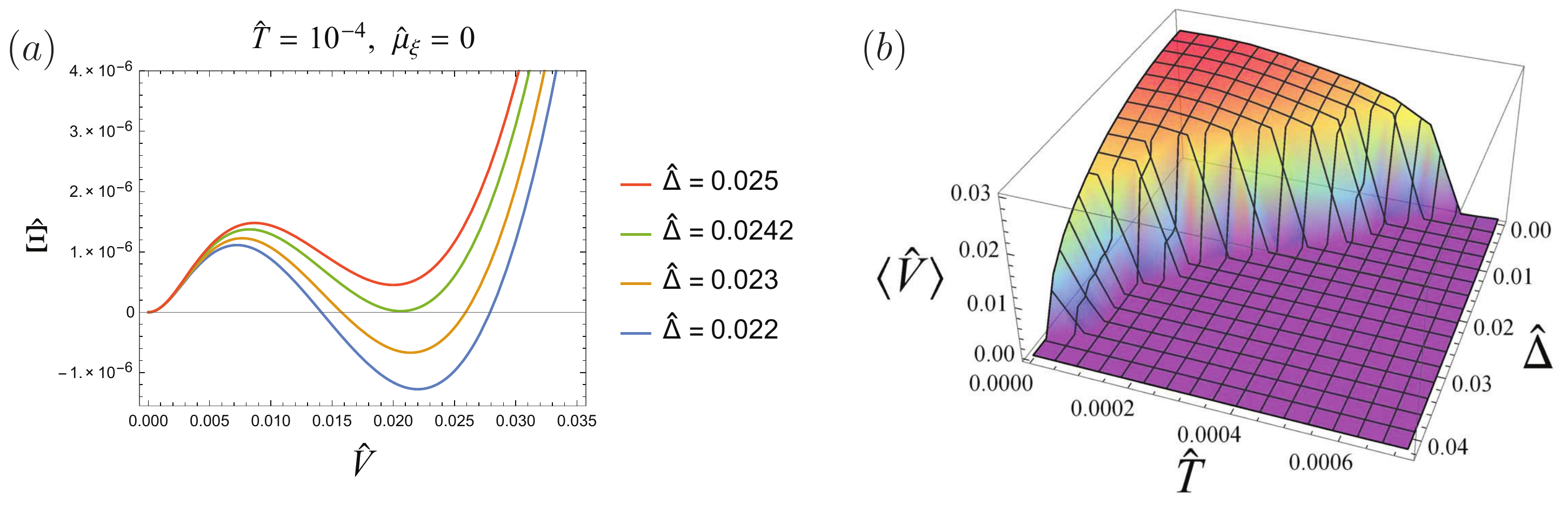}
	\vspace{-3pt}
	\caption{\label{fg:poten2}(a)~Thermodynamic potential $\hat{\Xi}$ and 
	(b)~condensate $\langle \hat{V} \rangle$ in the two-color two-flavor model  
	with $(\hat\mu,~\hat\mu_\xi,~\hat{G})=(0.5,~0,~5)$. 
	The phase transition is first order except when $\hat{T}=0$.}
\end{figure}

The thermodynamic potential at low temperature is plotted in Fig.~\ref{fg:poten2}(a). 
The tendency towards symmetry restoration is clearly visible for larger $\Delta$. 
%%%%%%%%%%%%%%%%%%%%%%%%%%%%
%%%%%%%%%%%%%%%%%%%%%%%%%%%%
%%%%%%%%%%%%%%%%%%%%%%%%%%%%
%%%%%%%%%%%%%%%%%%%%%%%%%%%%
In Fig.~\ref{fg:poten2}(b) the ($\hat{\Delta},~\hat{T}$)-dependence of the condensate 
is shown. We observe a clear first-order transition line separating the Kondo-screened phase and 
the free color moment phase.  
The condensate has a quite similar magnitude to Fig.~\ref{fg:phdiag3d}(a)
despite that the interaction here $\hat{G}=5$ is just half the coupling $\hat{G}=10$ 
in Fig.~\ref{fg:phdiag3d}(a). 
This is because the number of fermions that couple to the impurity is 
twice larger owing to the color degrees of freedom.  At any rate, the Kondo effect 
is substantially quenched by both $T$ and $\Delta$. In the present treatment 
the low-$T$ and low-$\Delta$ region is occupied by the ASC phase alone; 
one cannot see the OS-FP region depicted in Fig.~\ref{fg:OSp} because, not 
surprisingly, the mean-field approximation is too crude to describe 
the critical behavior of the intermediate fixed point. 

To gain insights into the Kondo problem in real quark matter, let us 
set the parameters of our model to those of the NJL model 
that are chosen so as to reproduce physical observables, such as 
the dynamical quark mass and the pion decay constant at $\mu=T=0$. 
($v$ is equal to the light velocity in QCD and need not be tuned. 
We neglect the small current quark masses in the following.) 
The quantities we need to extract from the literature are $G$, 
$\Lambda$ and $\mu$, as well as the dynamically determined value of the gap 
$\Delta$ at the given $\mu$ for $T=0$. We adopt the set $(G,\Lambda,\mu,\Delta)$ 
from Sec.~4.3.4 of \cite{Buballa:2003qv} (three-color two-flavor NJL model). 
The original parameters as well as those converted to dimensionless units 
are summarized below. The coupling $G/4$ in \eqref{eq:lagw} 
was equated to $g\;(=g^{\,}_{\rm E}=g^{\,}_{\rm M})$ in 
\cite[Sec.~4.3.4]{Buballa:2003qv}.  

%%%%%%%%%%%%%%%%%%%%%%%%%%%%%%%%%
%%%%%%%%%%%%%%%%%%%%%%%%%%%%%%%%%
\begin{center}
	\begin{tabular}{cccc}
		\hline \hline 
		\text{Quantities \cite{Buballa:2003qv}} & 
		$\begin{array}{c}
			\text{Quantities} \vspace{-4pt}
			\\
			\text{(in units of $\Lambda$)}
		\end{array}$ & &  
		\\\hline 
		~~
		$\begin{array}{l}
			\Delta  = 140\, [\text{MeV}] 
			\\
			\text{at}~\left\{
			\begin{array}{rl}
				\mu & = 500\, [\text{MeV}] 
				\\
				G & = 7.64\, [\text{GeV}^{-2}] 
				\\
				\Lambda & = 600\, [\text{MeV}]
			\end{array}\right.
		\end{array}$ & ~~$\begin{array}{l}
			\hat\Delta = 0.233 
			\\
			\text{at}~\left\{
			\begin{array}{rl}
				\hat\mu & = 0.833  
				\\
				\hat{G} & = 2.75
			\end{array} \right. 
		\end{array}$~~ & $\longrightarrow$ ~~
		& $\begin{array}{l}
			\hat\Delta^{\rm crit}= 0.0272 
			\\
			\text{at}~~\hat\mu_\xi = 0 
		\end{array}$~~
		\\\hline\hline 
	\end{tabular}
\end{center}
%%%%%%%%%%%%%%%%%%%%%%%%%%%%%%%%%
%%%%%%%%%%%%%%%%%%%%%%%%%%%%%%%%%

Using the values of $\hat\mu$ and $\hat{G}$ in the second column, 
we have calculated the critical gap $\hat\Delta^{\rm crit}$ in our model, 
beyond which the condensate $\langle\hat{V}\rangle$ vanishes. The result is given 
in the right column. The inequality 
$\hat\Delta > \hat\Delta^{\rm crit}$ implies that the Kondo screening 
is strongly suppressed by quark pairing at this density. Considering that the 
Kondo effect is suppressed in the high-density limit, too (Sec.~\ref{sc:os}), we conjecture 
that the range in which the QCD Kondo effect occurs is either non-existent or very narrow. 
Since our current understanding of the phase diagram of QCD is still far from complete, 
the analysis above only serves as a qualitative guide.

\subsection{\label{sc:sbs2}Shiba states \sc{sc:sbs2}}

Mid-gap states induced by an impurity in the 2SC phase can be analyzed by 
a suitable extension of the treatment in Sec.~\ref{sc:sbs1}.  
Let us consider a toy model of two-color two-flavor Dirac fermions with 
a classical impurity potential, 
\ba
	\hat{\cal H} & \equiv \int \dd^3x \ckakko{
		\psi^\dagger (
			iv \bm{\sigma\cdot} \nabla -\mu 
		) \psi 
		- \frac{\Delta}{2} \Big[\psi_\alpha 
		(\sigma^2 \tau^2 t^2)_{\alpha\beta} \psi_\beta + 
		\psi^{\dagger}_\alpha (\sigma^2 \tau^2 t^2)_{\alpha\beta} 
		\psi^\dagger_{\beta} \Big] 
		+ \psi^\dagger U(\mathbf{x}) \psi 
	}.
\ea
The impurity-free part of this Hamiltonian 
is invariant under $\SU(2)\times\SU(2)\times\SU(2)$ that acts on 
the spin($\sigma^i$)-flavor($\tau^i$)-color($t^i$) indices. 
In the Nambu basis $\Psi_{\bk}=\big(\psi_{\bk} \quad \psi^\dagger_{-\bk}\big)^{\rm T}$  
the Hamiltonian of a pure system reads  
\ba
	& \frac{1}{2}\int \frac{\dd^3 k}{(2\pi)^3}
	\npsi^\dagger_\bk
	\bep
		(-v \bk\bm{\cdot\sigma}-\mu)\otimes \1_4 
		& -\Delta \sigma^2 \tau^2 t^2
		\\
		-\Delta \sigma^2 \tau^2 t^2 & 
		(-v \bk\bm {\cdot \sigma}^{\rm T} + \mu)\otimes \1_4
	\eep
	\npsi_\bk 
	\equiv 
	\frac{1}{2}\int \frac{\dd^3 k}{(2\pi)^3}
	\npsi^\dagger_\bk
	\hat{\cal H}_\bk 
	\npsi_\bk \,, 
\ea
with $\1_4=\1_2^{\rm color}\otimes\1_2^{\rm flavor}$.  
The corresponding Green's function is given by
\ba
	\hat{G}_0(\bk,\omega) & \equiv \frac{1}{\omega-\hat{\cal H}_\bk}
	= \bep
		(\omega + v \bk\bm{\cdot\sigma} + \mu)\otimes \1_4 
		& \Delta \sigma^2 \tau^2 t^2
		\\
		\Delta \sigma^2 \tau^2 t^2 & 
		(\omega + v \bk\bm {\cdot \sigma}^{\rm T} - \mu)\otimes \1_4
	\eep^{-1}
	\\
	& = \frac{1}{\Y^2-4\mu^2v^2k^2}
	\begin{pmatrix}
		(\omega-\mu-v \bk\bm{\cdot \sigma})
		(\Y+2\mu v\bk\bm{\cdot \sigma}) \otimes \1_4 
		& 
		- \Delta \sigma^2 \tau^2 t^2 (\Y-2\mu v\bk\bm{\cdot \sigma}^{\rm T})
		\\
		- \Delta \sigma^2 \tau^2 t^2 (\Y+2\mu v\bk\bm{\cdot \sigma})  
		& 
		(\omega+\mu-v \bk\bm{\cdot \sigma}^{\rm T})
		(\Y-2\mu v\bk\bm{\cdot \sigma}^{\rm T}) \otimes \1_4 
	\end{pmatrix}
\ea
where $\Y = \omega^2-\mu^2-\Delta^2-v^2k^2$, as already defined below \eqref{eq:Gmatr}. 
All the terms linear in $\bk$ vanish after integration:
\ba
	\int \frac{\dd^3 k}{(2\pi)^3} 
	\hat{G}_0(\bk,\omega) 
	& = \int \frac{\dd^3 k}{(2\pi)^3} 
	\frac{1}{\Y^2-4\mu^2v^2k^2}
	\begin{pmatrix}
		[(\omega-\mu)\Y - 2 \mu v^2k^2] \1_8 & -\Delta \sigma^2\tau^2 t^2 \Y
		\\
		-\Delta \sigma^2\tau^2 t^2 \Y & 
		[(\omega+\mu)\Y + 2\mu v^2 k^2] \1_8
	\end{pmatrix}.
	\label{eq:Gzero2SC}
\ea
Evidently the above matrix structure is symmetric under 
the exchange of color/spin/flavor indices, implying that 
potentials proportional to $\sigma^3$ or $\tau^3$ or $t^3$ 
all lead to identical poles of the $T$-matrix \eqref{eq:ttt}. For instance, 
if a colorful potential [$U({\bf x})=u\,\delta({\bf x})t^3$] 
is given, one can readily diagonalize $\sigma^2\tau^2$ with a unitary 
transformation and obtain the same $T$-matrix 
as in Sec.~\ref{sc:sbs1}, which can be seen by juxtaposing 
\eqref{eq:Gzero2SC} with \eqref{eq:Gmatr}. Therefore the resulting 
energy levels of the intragap states coincide with \eqref{eq:omful}, 
but this time each level of \eqref{eq:omful} becomes four-fold degenerate, giving 
rise to 8 intragap states in total. This has a simple interpretation. Let us denote 
the flavor, color and spin of the quark at the impurity site by $(u,d)$, $(\Uparrow,\Downarrow)$ 
and $(\uparrow,\downarrow)$, respectively. Then, 
the four-fold degeneracy originates from the screening of the impurity $(\Uparrow)$ by one of   
$\big\{u^{\uparrow}_{\Downarrow}, u^{\downarrow}_{\Downarrow},  
d^{\uparrow}_{\Downarrow}, d^{\downarrow}_{\Downarrow}\big\}$. 
The first-order transition at intermediate coupling (Fig.~\ref{fg:levelcrossing}) 
means a transition from the LM phase to the ASC phase, both of which are four-fold degenerate. 

In the present analysis of a classical impurity, there is only one phase transition, beyond which  
the ASC phase remains the ground state for an arbitrarily large coupling. 
This is due to a limitation of the classical treatment. 
We know from the discussion in Sec.~\ref{sc:os} that the ground state at sufficiently strong coupling  
will be the SC phase in which all the four quarks $\big\{u^{\uparrow}_{\Downarrow}, 
u^{\downarrow}_{\Downarrow},  d^{\uparrow}_{\Downarrow}, 
d^{\downarrow}_{\Downarrow}\big\}$ participate in the screening of impurity's 
color ($\Uparrow$).  Moreover, between the ASC phase and the SC phase there can be two 
alternative screened phases in which two or three quarks couple to the impurity. 
Therefore we suspect that a heavy quark immersed in a color superconductor 
will experience more than one quantum phase transition if the coupling is swept 
from weak to strong coupling, or vice versa. 
A support to this conjecture comes from a recent finding \cite{Zitko2016} with NRG that 
the two- and three-channel Kondo model with a BCS gap undergoes 
multiple quantum phase transitions when the coupling is varied. 
In quark matter, of course the gauge coupling itself cannot be varied externally, 
but it runs dynamically as a function of density. Understanding the full evolution of 
mid-gap states in dense QCD by first-principles calculations is a challenging future problem.

For completeness let us also consider a more general potential 
$U=\delta(\mathbf{x}) (u_1 t^3 + u_2\sigma^3)$ with two independent couplings 
$u_1$ and $u_2$. In the limit $u_2\to 0$ it returns to the previous model 
$U({\bf x})=u\,\delta({\bf x})t^3$ with Shiba's spectrum \eqref{eq:omful}.  
The $T$-matrix poles follow from the equation
\ba
	0 & = \det\kkakko{ 
		\bep
			(u_1 t^3 + u_2 \sigma^3)\otimes \1_2^{\rm flavor} & 0 
			\\ 
			0 & (- u_1 t^3 - u_2 \sigma^3)\otimes \1_2^{\rm flavor}
		\eep^{-1} 
		- \int \frac{\dd^3 k}{(2\pi)^3} \hat{G}_0(\bk,\omega)
	} 
	\\
	& \simeq \det\kkakko{ 
		\frac{1}{u_1^2-u_2^2}\bep
			(u_1 t^3 - u_2 \sigma^3)\otimes \1_2^{\rm flavor} & 0 
			\\
			0 & (- u_1 t^3 + u_2 \sigma^3)\otimes \1_2^{\rm flavor}
		\eep 
		- 
		\frac{\pi \rho(\mu)}{2}\frac{1}{\sqrt{\Delta^2-\omega^2}}
		\bep
			-\omega \1_8 & \Delta \sigma^2 \tau^2 t^2 
			\\
			\Delta \sigma^2 \tau^2 t^2 & -\omega \1_8
		\eep
	},
	\label{eq:tmat2}
\ea
where in the last step we made the same approximations as for \eqref{eq:Tishiba}. 
%%%%%%%%%%%%%%%%%%%%%%%%%%%%%%%%%%%%%%
\begin{figure}[bt]
	\centering
	\includegraphics[width=11cm]{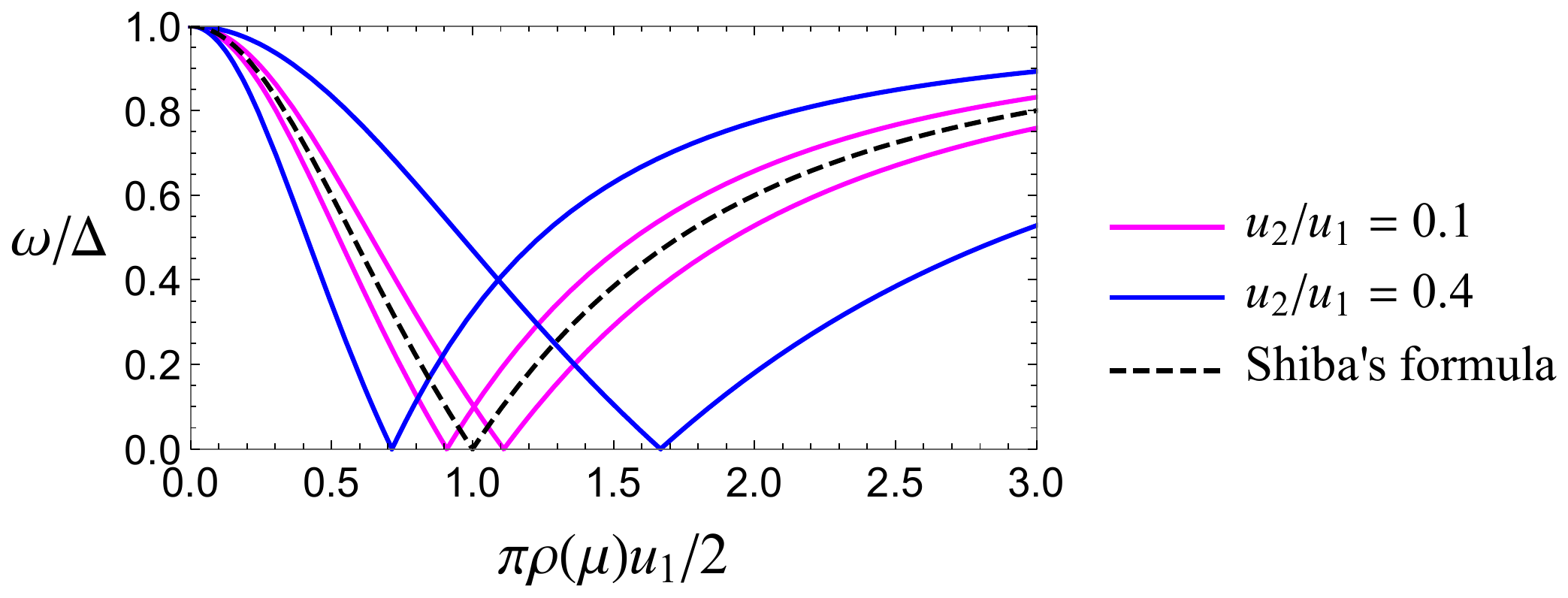}
	\vspace{-5pt}
	\caption{\label{fg:shiba2sc}
	Spectrum of mid-gap excited states for the potential 
	$U=\delta(\mathbf{x}) (u_1 t^3 + u_2\sigma^3)$ in comparison with 
	Shiba's formula \eqref{eq:omful}. Each level is doubly degenerate. 
	Negative energy levels (not shown) appear symmetrically about the horizontal axis.
	}
\end{figure}
%%%%%%%%%%%%%%%%%%%%%%%%%%%%%%%%%%%%%%
In Fig.~\ref{fg:shiba2sc} we plot the spectrum obtained by numerically solving \eqref{eq:tmat2}. 
The energy levels that were four-fold degenerate at $u_2=0$ now split into pairs of doubly 
degenerate levels at $u_2\ne 0$, due to the spin-symmetry breaking by the potential. 
For a fixed $u_2$ there are two level crossings. The first crossing at small $u_1$ is 
a transition from the LM phase to a flavor-doublet ASC phase 
in which the impurity $(\Uparrow\uparrow)$ is screened by either $u^{\downarrow}_{\Downarrow}$ 
or $d^{\downarrow}_{\Downarrow}$. This ASC remains the ground state beyond this 
transition. The second level crossing at larger $u_1$ occurs between the LM phase and 
another flavor-doublet ASC in which the impurity is screened by either 
$u^{\uparrow}_{\Downarrow}$ or $d^{\uparrow}_{\Downarrow}$. Since it is a level 
crossing between excited states, this does not correspond to a phase transition. 
The spectrum in Fig.~\ref{fg:shiba2sc} would be reliable at weak coupling but not 
at strong coupling due to the limitation of the current classical treatment; 
the SC phase is expected to emerge at strong coupling in a fully quantum treatment.

It should be noted that, contrastingly, no impurity-induced intragap state is found  
for a potential $U=u\delta(\mathbf{x})\sigma^3 t^3$. This is because 
$U$ does not act as a pair breaker: $U$ preserves the degeneracy of energies of 
$u^{\uparrow}_{\Downarrow}$ and $d^{\downarrow}_{\Uparrow}$ 
that make up a Cooper pair $(u^{\uparrow}_{\Downarrow}d^{\downarrow}_{\Uparrow})$. 
Therefore the Kondo effect is not caused by this potential however large $u$ is made. 
This is consistent with a perturbative analysis \cite{Kuramoto1998} 
showing that the interaction involving the double-exchange of spin and pseudospin  
does not grow by itself at low energy in the absence of other couplings. It is implied that 
the Kondo effect (ASC phase) found in the mean-field theory 
in Sec.~\ref{sc:sqk2} was actually an outcome of the two interactions in 
\eqref{eq:intint},
\ba
	\sum_{i=1}^{3} (\psi_f^\dagger \sigma^i \psi_f^{})(\xi^\dagger \sigma^i \xi)  
	\quad \text{and} \quad 
	\sum_{j=1}^{3} (\psi_f^\dagger t^j \psi_f^{})(\xi^\dagger t^j \xi) \,,  
	%\qquad \quad [\text{cf.~\eqref{eq:intint}}]
\ea
with the other two pieces $(\psi_f^\dagger\psi_f^{})(\xi^\dagger\xi)$ and 
$\sum_{i,j=1}^{3}(\psi_f^\dagger \sigma^i t^j \psi_f^{})(\xi^\dagger \sigma^i t^j \xi)$ 
playing no decisive role.

\section{\label{sc:cl}Conclusions \sc{sc:cl}}
In this paper, we have studied the interplay between superconductivity and the Kondo effect
in emergent and intrinsic relativistic systems.
In the absence of superconductivity, the overscreened Kondo effect prevails in such systems
due to the multichannel nature originating in the various degrees of freedom.
Once a bulk superconducting gap shows up, we show in terms of
the scaling and the existing NRG analyses that
in addition to the OS-FP and LM phases,
the ASC phase in which a minimal number of conduction fermions
participates in screening 
is also allowed in the phase diagram
in the weak-coupling limit. In the strong-coupling limit, the LM phase
should be replaced by the SC phase in which all channels couple to
an impurity in a symmetric manner.

We also performed the slave-boson mean-field analysis in two relativistic models.
While this analysis cannot distinguish the OS-FP phase from
the ASC phase, the transition between the Kondo and local moment phases
can be directly captured by the presence or absence of a nonvanishing order parameter $V$. 
This way we showed the suppression of the Kondo screening at finite temperature and a pairing gap.

We pointed out that the QCD Kondo effect discussed in the high-density limit is
suppressed by color superconductivity, since $\Delta\gg \Lambda_{\text{K}}$.
Thus, the phase realized in this limit may be the LM one without the Kondo screening.
At the same time, we demonstrated with the $T$-matrix and classical impurity methods
that the Shiba states localized in the vicinity of an impurity
are allowed in the LM phase. Thus, the excited states in color superconductors 
become nontrivial in the presence of an impurity.
A future problem is to discuss a physical influence due to the Shiba state in color superconductivity.
An unbiased analysis beyond the classical spin method is needed to directly show 
quantum phase transitions allowed for an impurity with multiple quantum numbers.

\begin{acknowledgments}
	T.~K. was supported by the RIKEN iTHES project.  
	T.~K. thanks Rok $\check{\rm Z}$itko for a valuable correspondence 
	concerning \cite{Zitko2016}. 
\end{acknowledgments}

\appendix* 
\section{\label{sc:ap}Scaling function to two loops}

This appendix provides a brief derivation of scaling functions 
for the models \eqref{eq:mdl} and \eqref{eq:mdl2} in Sec.~\ref{sc:os}.  
The essential idea of Anderson's scaling analysis \cite{Anderson1970} was that the effect of 
integrating out high-energy modes can be compensated by modifying parameters 
in the original model. Here we shall follow the same idea by introducing an IR cutoff around 
the Fermi surface, whose variation can be absorbed by redefinition of the coupling. 
To renormalize the model \eqref{eq:mdl} we introduce a bare coupling $G_{\B}$ and a 
bare impurity field $\xi_{\B}$ that are independent of the IR cutoff:
\ba
	\mathcal{L} 
	& = \sum_{f=1}^{N_f}\sum_{a=1}^{N_c}\psi^\dag_{fa} (\der_\tau-\mu+i \bm{\sigma\cdot}\nabla)\psi_{fa}  
	+ \sum_{a=1}^{N_c} \xi^\dag_{\B a} (\der_\tau-\mu_\xi)\xi_{\B a}^{} 
	+ G_{\B} \sum_{f=1}^{N_f} \sum_{a,b=1}^{N_c} 
	\psi_{fa}^\dag \psi^{}_{fb} \xi_{\B b}^\dag \xi_{\B a}^{} \,. 
	\label{eq:mdlAPP}
\ea

%%%%%%%%%%%%%%%%%%%%%%%%%%%%%%%
%%%%%%%%%%%%%%%%%%%%%%%%%%%%%%%
\begin{figure}[t]
	\centering
	\includegraphics[height=.18\textwidth]{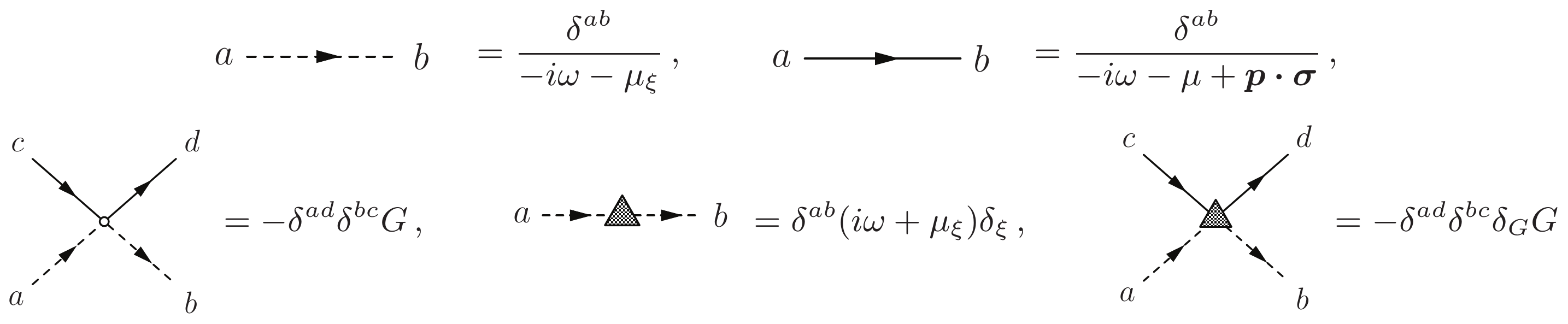}
%	\raisebox{-1.5mm}{\includegraphics[width=22mm]{fg_diag_1}}
%	~~$\displaystyle= \frac{\delta^{ab}}{-i\omega-\mu_\xi}$\,, \qquad 
%	\raisebox{-1.5mm}{\includegraphics[width=22mm]{fg_diag_2}}
%	~~$\displaystyle=\frac{\delta^{ab}}{-i\omega-\mu+\bm{p\cdot\sigma}}$\,, 
%	\vspace{5pt}
%	\\
%	\raisebox{-8mm}{\includegraphics[width=20mm]{fg_diag_3}}
%	$=-\delta^{ad}\delta^{bc}G$\,, \qquad 
%	\raisebox{-1mm}{\includegraphics[width=22mm]{fg_diag_c1}}
%	$=\delta^{ab}(i\omega+\mu_\xi)\delta_\xi$\,, \qquad 
%	\raisebox{-8mm}{\includegraphics[width=20mm]{fg_diag_c2}}
%	$=-\delta^{ad}\delta^{bc}\delta_{G}G$
	%%%
	%%%
	%\vspace{-5pt}
	\caption{\label{fg:Frules}
	Feynman rules for the model \eqref{eq:mdlren}. The last two vertices are counterterms.}
	\vspace{20pt}
	\centering 
	\includegraphics[height=.1\textwidth]{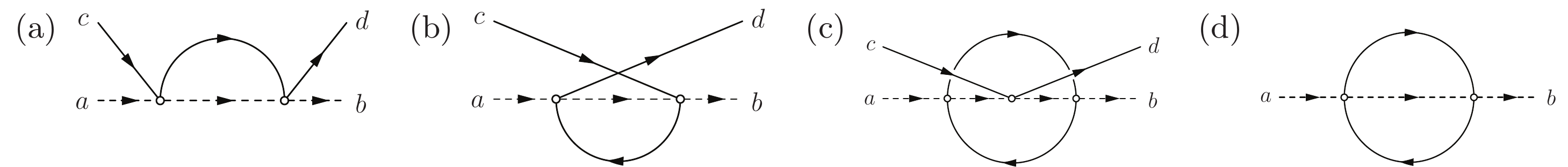}
%	\raisebox{13mm}{(a)} \raisebox{5mm}{\includegraphics[width=3cm]{fg_diag_4}}~~
%	\raisebox{13mm}{(b)} \includegraphics[width=3cm]{fg_diag_5}~~
%	\raisebox{13mm}{(c)} \includegraphics[width=3cm]{fg_diag_6}~~
%	\raisebox{13mm}{(d)} \includegraphics[width=3cm]{fg_diag_7}~
	\vspace{2pt}
	\caption{\label{fg:diags}Feynman diagrams contributing to the $\beta$ function of $G$ 
	at $\calO(G^3)$. The diagram (c) is the leading one at $N_f\gg 1$. 
	}
\end{figure}
%%%%%%%%%%%%%%%%%%%%%%%%%%%%%%%
%%%%%%%%%%%%%%%%%%%%%%%%%%%%%%%

Let us introduce a renormalized coupling $G$ and a renormalized field $\xi$,
\ba
	\xi_{\B} = \sqrt{Z_\xi}\,\xi\,, \quad G_\B=Z_\xi^{-1}Z_G G\,, 
	\quad \delta_\xi \equiv Z_\xi -1\,, \quad \delta_G \equiv Z_G-1\,,
\ea
in terms of which the Lagrangian is cast into the form
\ba
	\mathcal{L} & = 
	\psi^\dag_{fa} (\der_\tau-\mu+i \bm{\sigma\cdot}\nabla)\psi_{fa}  
	+ \xi_a^\dag (\der_\tau-\mu_\xi)\xi_a 
	+ G \psi_{fa}^\dag \psi^{}_{fb} \xi_{b}^\dag \xi_{a}^{} 
	+ \delta_{\xi} \xi_a^{\dag} (\der_\tau-\mu_\xi)\xi_a  
	+ \delta_{G} G \psi_{fa}^\dag \psi^{}_{fb} \xi_{b}^\dag \xi_{a}^{} 
	\label{eq:mdlren}
\ea
where the summation over repeated indices is assumed. 
The resulting Feynman rules are summarized in Fig.~\ref{fg:Frules} and 
relevant diagrams for the scaling of $G$ at $N_f\gg 1$ are listed in Fig.~\ref{fg:diags}. 
Because Fig.~\ref{fg:diags}(a) $\propto 
\delta^{ap}\delta^{cq}\delta^{pb}\delta^{qd}=\delta^{ab}\delta^{cd}$, it 
is a potential scattering $\sim \psi_{fa}^\dagger \psi_{fa} \xi_b^\dagger \xi_b$ 
and does not contribute to the renormalization of $G$. 
The same holds for Fig.~\ref{fg:diags}(c), since 
$\text{Fig.~\ref{fg:diags}(c)} \propto \delta^{ap}\delta^{qr} \cdot 
\delta^{rd}\delta^{cs} \cdot \delta^{pb}\delta^{sq} = \delta^{ab}\delta^{cd}$\,. 
Let us consider Fig.~\ref{fg:diags}(b). The amplitude is straightforwardly obtained as
\ba
	\delta^{ad}\delta^{pq}\cdot \delta^{bc}\delta^{pq}\cdot 
	G^2 \int \frac{\dd^3 k}{(2\pi)^3}\int \frac{\dd k_0}{2\pi}
	\frac{1}{-ik_0-\mu_\xi} \frac{1}{-i(k_0+\omega)-\mu+\bm{k \cdot \sigma}}\,. 
\ea
One can use the relation 
\ba
	\frac{1}{-i(k_0+\omega)-\mu+\bm{k \cdot \sigma}} 
	& = \frac{1}{2}\kkakko{
		\frac{1}{-i(k_0+\omega)-\mu+|\bm{k}|} 
		+ \frac{1}{-i(k_0+\omega)-\mu-|\bm{k}|}
	} \1_2
	- \frac{\bm{k \cdot \sigma}}{[-i(k_0+\omega)-\mu]^2-\bm{k}^2}\,,
\ea
in which the last term odd in $\bm{k}$ vanishes after integration. The first (second) term 
in the square bracket represents the particle (antiparticle) contribution. Then 
\ba
	\text{Fig.~\ref{fg:diags}(b)} & = [\text{Fig.~\ref{fg:diags}(b)}]^{\rm (\text{p})} 
	+ [\text{Fig.~\ref{fg:diags}(b)}]^{\rm (\text{ap})} \,,
	\\
	[\text{Fig.~\ref{fg:diags}(b)}]^{\rm (\text{p})} 
	& = \frac{1}{2}\delta^{ad}\delta^{bc} N_c G^2 
	\int \frac{\dd^3 k}{(2\pi)^3}\int \frac{\dd k_0}{2\pi}
	\frac{1}{-ik_0-\mu_\xi} \frac{1}{-i(k_0+\omega)-\mu+|\bm{k}|} 
	\\
	& = \frac{1}{2}\delta^{ad}\delta^{bc} N_c G^2 
	\int \frac{\dd^3 k}{(2\pi)^3}
	\frac{\theta(\mu-|\bm{k}|)}{-i\omega-\mu+|\bm{k}|+\mu_\xi}
	\\
	& = \frac{1}{2}\delta^{ad}\delta^{bc} N_c G^2 
	\int_{0}^{\mu-D} \frac{\dd k\;k^2}{2\pi^2} \frac{1}{-\mu+k}
	\label{eq:2bi}
	\\
	& = \frac{1}{2}\delta^{ad}\delta^{bc} N_c G^2 
	\rho \, [\log (D/\mu) + \calO(1)] \,, 
\ea
where we used $\mu_\xi\leq 0$ in performing the contour integration over $k_0$. 
To derive \eqref{eq:2bi} we continued $\omega$ to real frequency 
($i\omega\to \omega+i\epsilon$) and assumed that 
all external particles are on-shell, so that $-i\omega+\mu_\xi=0$. The resulting integral is 
divergent due to the contribution from the Fermi surface $k\approx \mu$, which is regularized by a cutoff $D$.   
The density of states in the normal phase at the Fermi surface is denoted by 
$\displaystyle \rho \equiv \frac{4\pi\mu^2}{(2\pi)^3} = \frac{\mu^2}{2\pi^2}$. 
The logarithmic divergence in $D$ is absorbed by a counterterm 
\ba
	\delta_G & = \frac{1}{2} N_c \GG \log (D/\mu) \qquad 
	\text{with}\quad \GG\equiv \rho G\,.
	\label{eq:deltaG}
\ea
On the other hand the antiparticle contribution
\ba
	[\text{Fig.~\ref{fg:diags}(b)}]^{\rm (\text{ap})} 
	& = \frac{1}{2}\delta^{ad}\delta^{bc} N_c G^2 
	\int \frac{\dd^3 k}{(2\pi)^3}\int \frac{\dd k_0}{2\pi}
	\frac{1}{-ik_0-\mu_\xi} \frac{1}{-i(k_0+\omega)-\mu-|\bm{k}|} 
\ea
is free from the IR singularity at the Fermi surface and is irrelevant to 
the Kondo effect.

Next we consider Fig.~\ref{fg:diags}(d), which contributes to the 
wave function renormalization of impurity. Recalling that there are 
$N_f$ flavors of fermions circulating around the loop, and attaching 
$(-1)$ for a single loop, we obtain the amplitude
\ba
	& (-1)N_f \cdot \delta^{ap}\delta^{cq}
	\cdot \delta^{pb}\delta^{cq}\cdot G^2 
	\int\frac{\dd^4 p}{(2\pi)^4} \int\frac{\dd^4 k}{(2\pi)^4}
	\frac{1}{-i(\omega-k_0+p_0)-\mu_\xi}\tr\mkakko{
		\frac{1}{-ip_0-\mu+\bm{p\cdot \sigma}}
		\frac{1}{-ik_0-\mu+\bm{k\cdot \sigma}}
	}
	\notag
	\\
	= \; & - \delta^{ab}N_cN_fG^2 \cdot\mkakko{\frac{1}{2}}^2\tr(\1_2)\cdot
	\int\frac{\dd^4 p}{(2\pi)^4} \int\frac{\dd^4 k}{(2\pi)^4}
	\frac{1}{-i(\omega-k_0+p_0)-\mu_\xi}
	\bigg( \frac{1}{-ip_0-\mu+|\bm{p}|}\frac{1}{-ik_0-\mu+|\bm{k}|}
	\notag
	\\
	& \qquad 
	+\frac{1}{-ip_0-\mu-|\bm{p}|}\frac{1}{-ik_0-\mu+|\bm{k}|}
	+\frac{1}{-ip_0-\mu+|\bm{p}|}\frac{1}{-ik_0-\mu-|\bm{k}|}
	+\frac{1}{-ip_0-\mu-|\bm{p}|}\frac{1}{-ik_0-\mu-|\bm{k}|}
	\bigg)
	\\
	=\; & \frac{1}{2}\delta^{ab}N_cN_fG^2
	\int\frac{\dd^3 p}{(2\pi)^3} \int\frac{\dd^3 k}{(2\pi)^3}
	\theta(|\bm{k}|-\mu)
	\bigg[
		\frac{\theta(\mu-|\bm{p}|)}
		{-i\omega-\mu_\xi-|\bm{p}|+|\bm{k}|}
		+ 
		\frac{1}{-i\omega-\mu_\xi+|\bm{p}|+|\bm{k}|}
	\bigg]\,.
\ea
The second term in the square bracket has no IR singularity and is irrelevant to the Kondo effect. 
To determine the wave function renormalization we need to compute the derivative with respect to $\omega$, 
\ba
	& \lim_{i\omega+\mu_\xi\to 0}\frac{\der}{\der (i\omega)} 
	\bigg[
		\frac{1}{2}\delta^{ab}N_cN_fG^2
		\int\frac{\dd^3 p}{(2\pi)^3} \int\frac{\dd^3 k}{(2\pi)^3}
		\frac{\theta(|\bm{k}|-\mu)\theta(\mu-|\bm{p}|)}
		{-i\omega-\mu_\xi-|\bm{p}|+|\bm{k}|}
	\bigg]
	\notag 
	\\
	= \; & \frac{1}{2}\delta^{ab}N_cN_fG^2
		\int\frac{\dd^3 p}{(2\pi)^3} \int\frac{\dd^3 k}{(2\pi)^3}
		\frac{\theta(|\bm{k}|-\mu)\theta(\mu-|\bm{p}|)}
		{(-|\bm{p}|+|\bm{k}|)^2}
	\\
	= \; & \frac{1}{2}\delta^{ab}N_cN_fG^2 
		\frac{4\pi}{(2\pi)^3}\frac{4\pi}{(2\pi)^3}
		\int_0^{\mu-D} \!\!\!\!\!\!\! \dd p\; p^2 \int_{\mu+D}^\Lambda \!\!\! \dd k\; k^2 
		\frac{1}{(-|\bm{p}|+|\bm{k}|)^2}
%	\\
%	=\; & \frac{1}{2}\delta^{ab}N_cN_fG^2 
%		\frac{4\pi}{(2\pi)^3}\frac{4\pi}{(2\pi)^3}
%		[ -\mu^4 \log (D/\mu) + \calO(1) ]
	\\
	= \; & - \frac{1}{2}\delta^{ab}N_cN_f \GG^2\, [\log (D/\mu) + \calO(1)]\,,
\ea
where $D$ cuts off the IR divergence and $\Lambda$ the UV divergence. 
(The coefficient of $\log D$ in the last line does not depend on details of 
the way the IR cutoff is imposed.) The resulting IR divergence is absorbed by 
the counterterm
\ba
	\delta_\xi & = \frac{1}{2}N_cN_f \GG^2 \log (D/\mu) \,. 
	\label{eq:deltaxi}
\ea
The $\beta$ function for the dimensionless coupling $\GG\equiv \rho G$ 
can now be computed. Recalling that the bare coupling $G_\B$ 
does not depend on $D$, one obtains \eqref{eq:betaI} from \eqref{eq:deltaG} and \eqref{eq:deltaxi}. 

The scaling function for the second model \eqref{eq:mdl2} can be obtained from that of the first model by 
just letting $N_c=2$, with the caveat that one must divide the contribution of Fig.~\ref{fg:diags}(d) 
by 2 to avoid double counting of color and spin. This yields \eqref{eq:betaII}.

\bibliographystyle{apsrev4-1}
%\bibliography{Shiba_refs}
\bibliography{Shiba_Dirac_PRD_v2.bbl}
\end{document}